\documentclass[aps,twocolumn,showpacs]{revtex4}
\usepackage{dcolumn}
\usepackage{graphicx}
\usepackage{amsmath}
\usepackage{amsfonts}
\usepackage{amssymb}
\usepackage{psfrag}
\usepackage{wrapfig}
\usepackage{subfigure}
\usepackage{makeidx}
\usepackage{bm}
\usepackage{epsf}

\begin{document}

\title{ Rogue Wave Solutions of a Three-Component Coupled Nonlinear Schr\"{o}dinger Equation}
\author{Li-Chen Zhao$^{1,2}$, Jie Liu$^{1,3}$}\email{liu_jie@iapcm.ac.cn}
\address{$^1$Science and Technology Computation Physics Laboratory,
 Institute of Applied Physics and Computational Mathematics, Beijing 100088,
China}
\address{$^2$Department of Modern Physics, University of Science and Technology of China, Hefei 230026, China}
\address{$^3$Center for Applied Physics and Technology, Peking University, Beijing 100084,
China}
\date{\today}
\begin{abstract}
We investigate rogue wave solutions in a three-component coupled
nonlinear Schr\"{o}dinger equation. With the certain requirements on
the backgrounds of components, we construct a new multi-rogue wave
solution that exhibits a structure like a four-petaled flower in
temporal-spatial distribution, in contrast to the eye-shaped
structure in one-component or two-component systems. The results
could be of interest in such diverse fields as Bose-Einstein
condensate, nonlinear fibers and super fluid.
\end{abstract}
 \maketitle

\section{Introduction}
Rogue wave(RW) is localized in both space and time, which seems to
appear from nowhere and disappear without a trace
\cite{N.Akhmediev,C. Kharif}. It is one of the most fascinating
phenomena in nature and has been observed recently in nonlinear
optics \cite{D.R. Solli} and water wave tank \cite{A. Chabchoub}.
The studies of RW in single-component system have
 indicated that the rational solution of the nonlinear Schr\"{o}dinger equation
(NLS) can be used to describe the phenomenon well \cite{Ankiewicz,R.
Osborne,V. Voronovich,Akhmediev}.

A variety of complex systems, such as Bose-Einstein condensates,
nonlinear optical fibers, etc., usually involve more than one
component \cite{Baronio}. Recent studies are extended to
 RWs in two-component systems
\cite{Baronio,Ling2,Bludov,zhao2}. Some new structures such as dark
RW have been presented numerically \cite{Bludov} and analytically
\cite{zhao2}. Moreover, it was found that two RWs can emerge in the
coupled system, which are quite distinct from high-order RW in
one-component system \cite{zhao2}. In the two-component coupled
systems, the interaction between RW and other nonlinear waves is
also a hot topic of great interest \cite{Baronio,Ling2,zhao2}. It
was shown that RW attracts dark-bright wave  in \cite{Baronio}.

In the present paper, we further extend to investigate a
three-component coupled system considering the number of the modes
coupled in the complex systems is usually more than two. With the
certain requirements on the backgrounds of components, we construct
a new rational solution that can be used to describe the dynamics of
one RW, two RWs, and three RWs in the system. One structure like a
four-petaled flower  is found in the coupled system: there are two
humps and two valleys around a center in the temporal-spatial
distribution, which is quite distinct from well-known eye-shaped one
presented before. We discuss the possibilities to observe them in
nonlinear fibers.

The paper is organized as follows. In Section {\rm II}, we present
 exact vector RW solutions and the explicit conditions under which
 they could exist. The dynamics of them are discussed in detail. In
Section {\rm III}, the possibilities to observe them are discussed.
The conclusion and discussion are made in Section {\rm IV}.

\section{ Exact vector rogue wave solutions and their dynamics}
It is well known that coupled NLS equations are often used to
describe the interaction among the modes in nonlinear optics,
components in BEC, etc.. We begin with the well-known
three-component coupled NLS, which can be written as
\begin{eqnarray}
 &&i\frac{\partial \psi_1}{\partial t}+\frac{\partial^2
\psi_1}{\partial x^2}+2[|\psi_1|^2+|\psi_2|^2+|\psi_3|^2]\psi_1 =0,\nonumber\\
&&i\frac{\partial \psi_2}{\partial t}+\frac{\partial^2
\psi_2}{\partial x^2}+2[|\psi_1|^2+|\psi_2|^2+|\psi_3|^2]\psi_2=0,\nonumber\\
&&i\frac{\partial \psi_3}{\partial t}+\frac{\partial^2
\psi_3}{\partial x^2}+2[|\psi_1|^2+|\psi_2|^2+|\psi_3|^2]\psi_3=0,
\end{eqnarray}
where $\psi_j$ (j=1,2,3) represent the wave envelopes, $t$ is the
evolution variable, and $x$ is a second independent variable. The
Eq.(1) has been solved to get vector soliton solution on trivial
background through Horita bilinear method in \cite{Lakshmanan}.
Performing Darboux-transformation from a trivial seed soltuion with
$\psi_3=0$, one could get the bright-bright solitons in \cite{Zhao}.
It has been reported that solitons could collide inelastically and
there are shape-changing collisions for coupled system, which are
different from uncoupled system \cite{Lakshmanan}. However, it is
not possible to study vector RW on trivial background. Next, we will
solve it from nontrivial seed solutions. The nontrivial seed
solutions are derived as follows
\begin{eqnarray}
\psi_{10} &=&s_1 \exp{[i 2( s_1^2+
s_2^2+s_3^2)t+ik_1 x-i k_1^2 t]},\\
\psi_{20}&=&s_2 \exp{[i 2(s_1^2+s_2^2+s_3^2)t+ik_2 x-i k_2^2 t]},\\
\psi_{30}&=&s_3 \exp{[i 2(s_1^2+ s_2^2+s_3^2)t+ik_3 x-i k_3^2 t]}
\end{eqnarray}
where $s_j$(j=1,2,3) are arbitrary real constants, and denote the
backgrounds in which localized nonlinear waves emerge. $k_1$, $k_2$
and $k_3$ denote the wave vectors of the plane wave background in
the three components respectively. From physical viewpoint, the
relative wave vector value is important. One of the three components
can be seen as a reference to define the wave vectors of the other
two. Then, we can set $k_2=0$ without losing generality. To derive
the rational solutions, we find that there are some requirements on
the amplitude of each component, and the difference of their wave
vectors should be related with the amplitude in a certain way. The
conditions under which one can get vector RW with no other type
waves are
\begin{eqnarray}
&&k_1=-k_3, k_3=\frac{\sqrt{2}}{2}s_1,\\
&&s_2= \frac{\sqrt{2}}{2}s_1, s_3=s_1.
\end{eqnarray}
With the conditions and $s_1=1$, the generic form of vector RWs
could be given as
\begin{eqnarray}
\psi_1&=&\left(1 - \frac{H_1(x,t)}{G_1(x,t)}\right)\exp{[i\frac{9 t}{2} -i\frac{x}{\sqrt{2}}]},\\
\psi_2&=&\left(1 -
\frac{H_2(x,t)}{G_2(x,t)}\right)\frac{\exp{[5 i t]}}{\sqrt{2}},\\
\psi_3&=&\left(1 - \frac{H_3(x,t)}{G_3(x,t)}\right) \exp{[i\frac{9
t}{2}+i\frac{x}{\sqrt{2}}]},
\end{eqnarray}
The expressions for $H_j$ and $G_j$ are given in the Appendix part.
It is seen that they are all rational forms. Between the
expressions, $A_j$ (j=1,2,3,4) are arbitrary real parameters.
Therefore, the vector waves solution could be vector RWs solution,
which can be verified by the following RWs plots. There are mainly
three cases for the generalized vector RW solution.

\begin{figure}[htb]
\centering
\subfigure[]{\includegraphics[height=35mm,width=40mm]{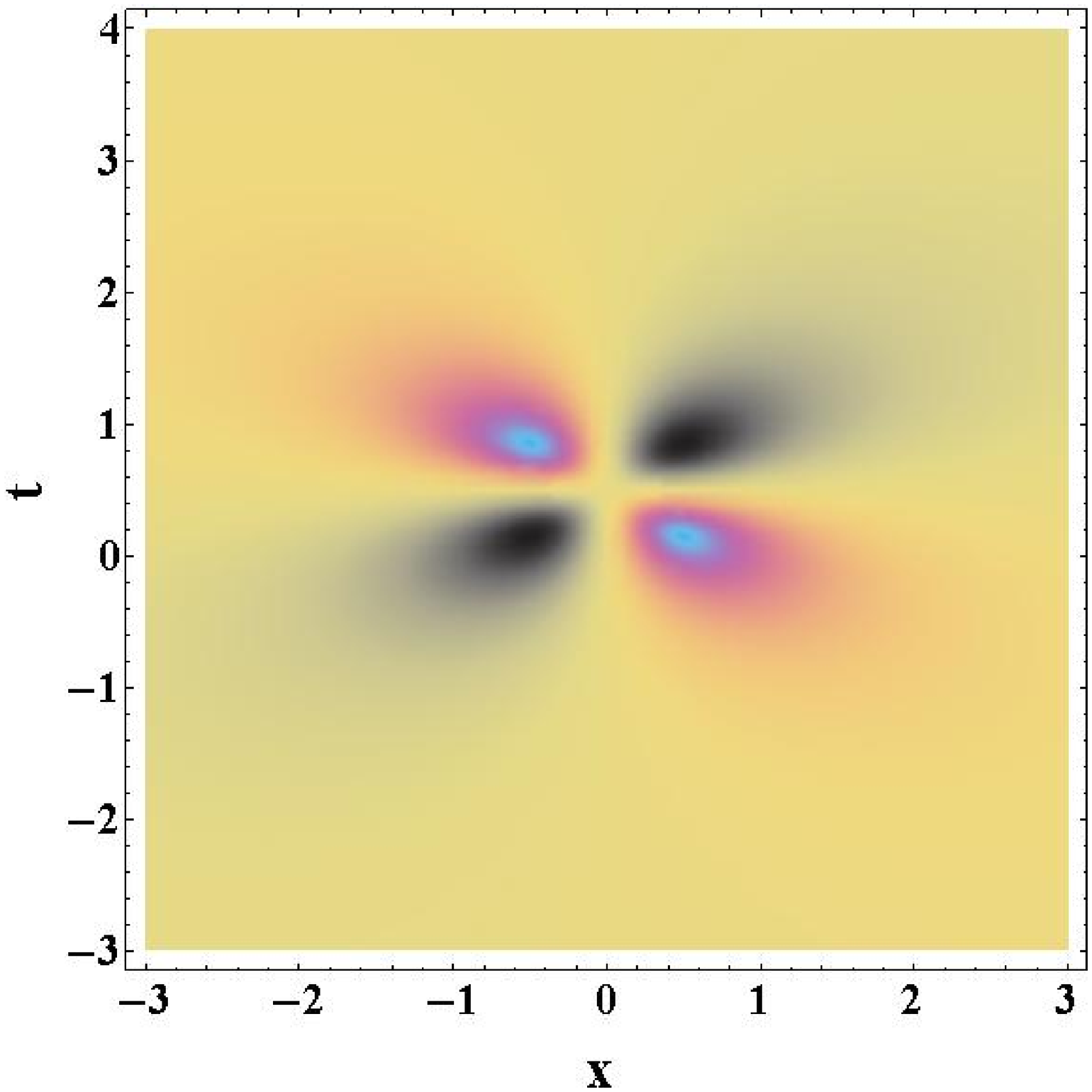}}
\hfil
\subfigure[]{\includegraphics[height=36mm,width=45mm]{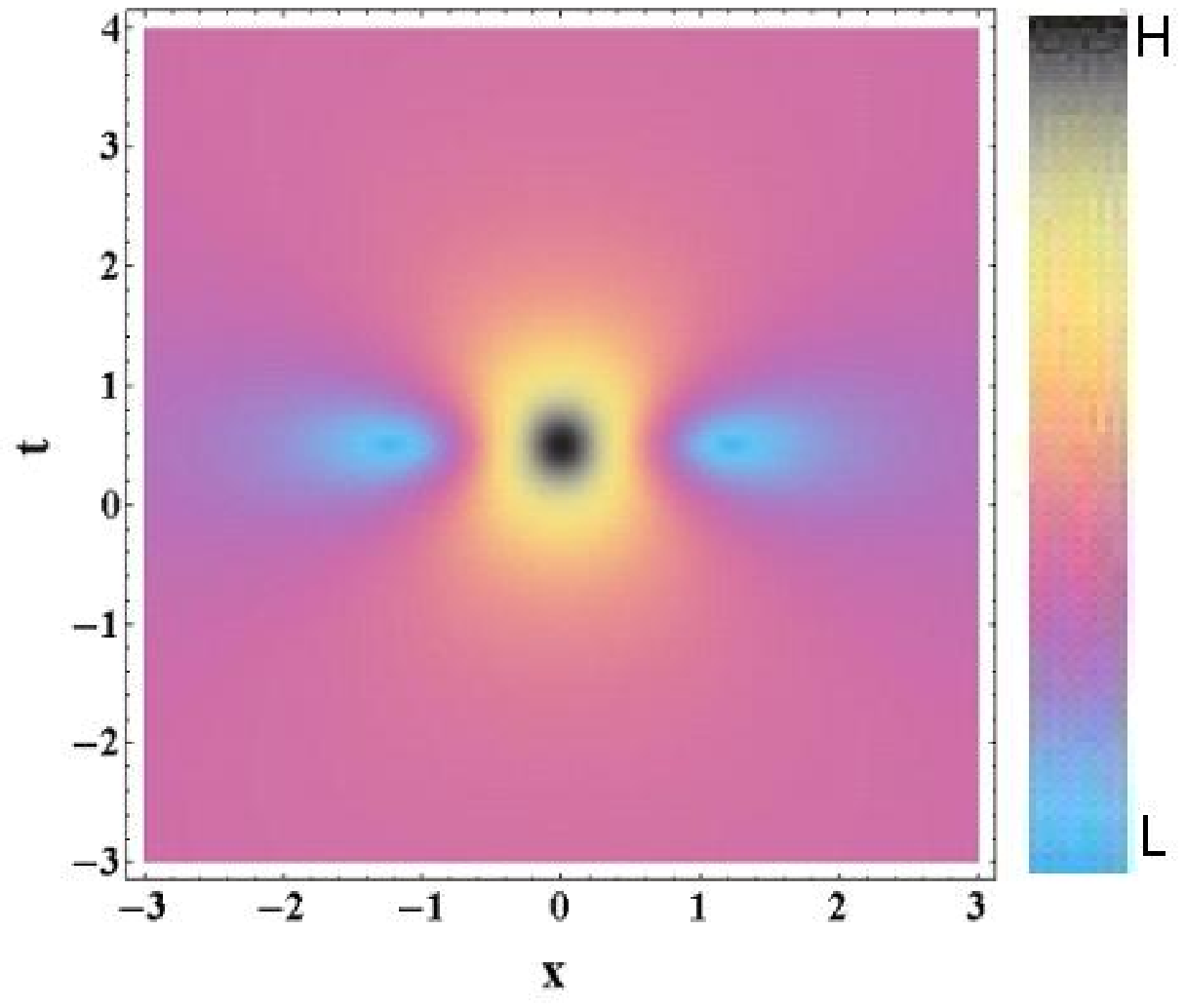}}
\hfil
\subfigure[]{\includegraphics[height=35mm,width=40mm]{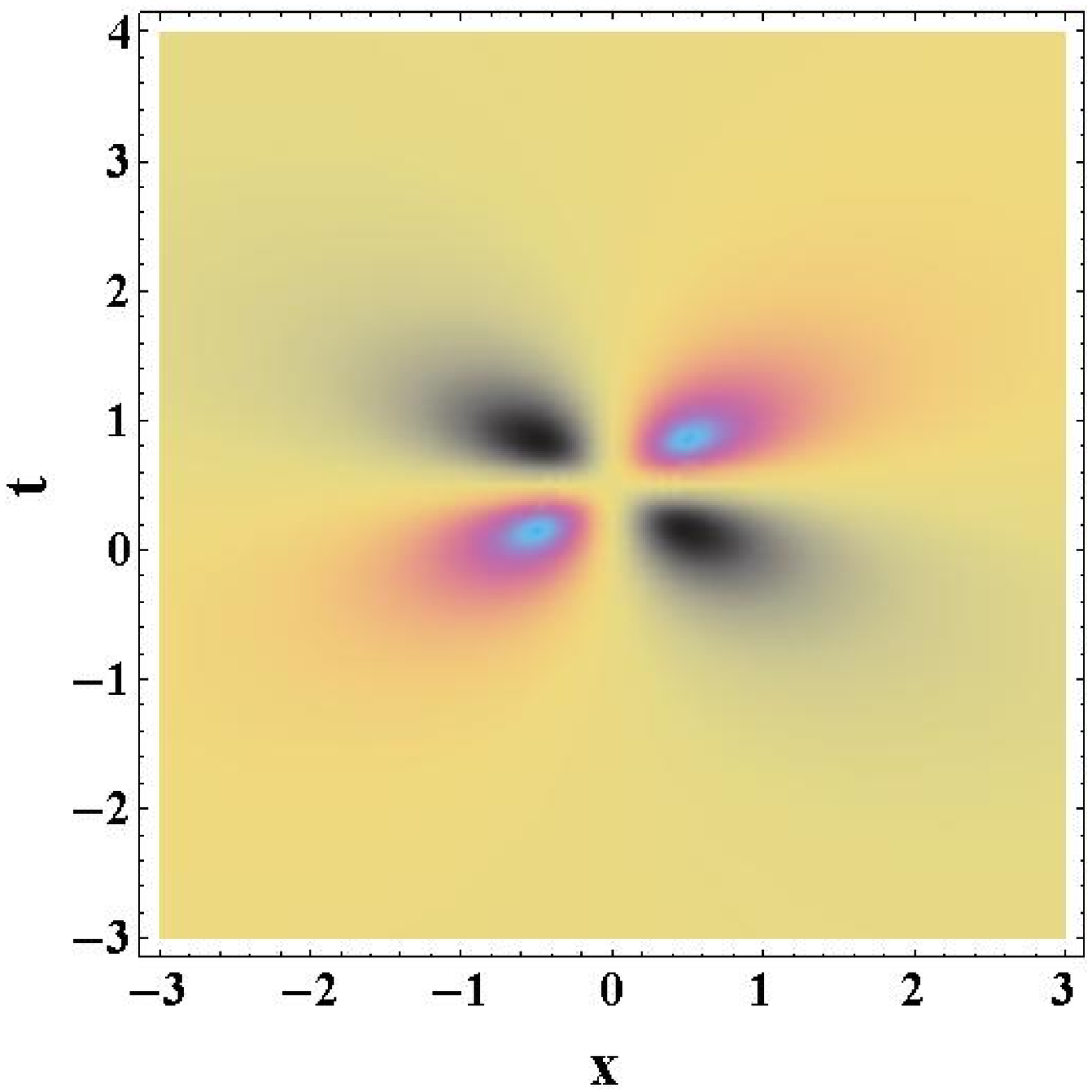}}
\hfil
\subfigure[]{\includegraphics[height=36mm,width=45mm]{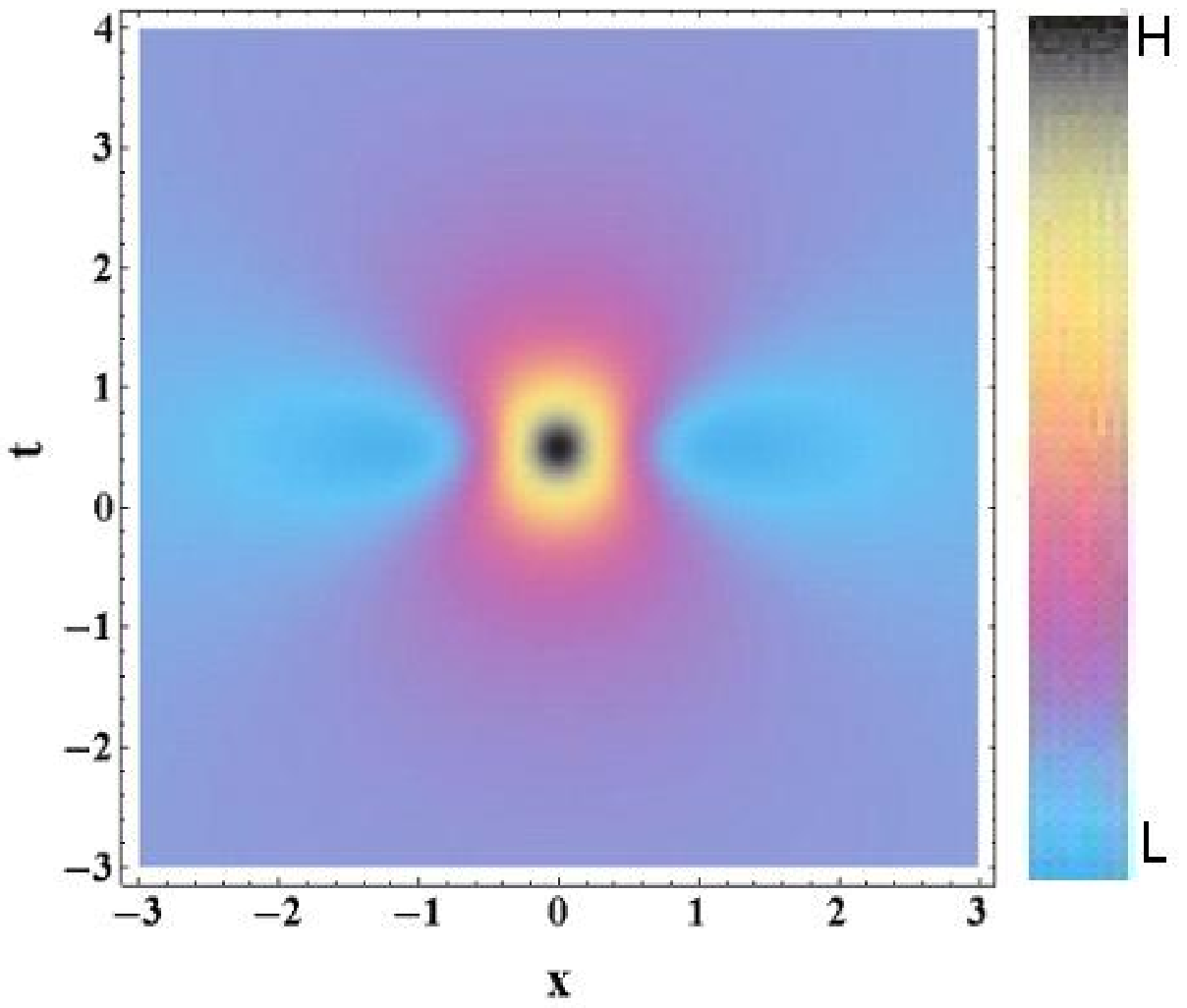}}
\caption{(color online)  The evolution plot of one RW in coupled
system, (a) for one RW in $\psi_1$ component, (b) for one RW in
$\psi_2$ component, (c) for one RW in $\psi_3$ component. (d) for
the whole density of the three components. It is seen that the
density distribution shapes of RW in $\psi_1$ and $\psi_3$ are
different from the eyes shape. The coefficients are $A_1 = 0, A_2 =
1, A_3 = 0$, and $ A_4 = 0$. The "L" and "S" in color bar denote
large and small values of density respectively. This holds for all
figures. }
\end{figure}

 \emph{One vector rouge wave}---
When $A_3=0, A_4=0$, only one vector RW can be observed in all
components, as shown in Fig. 1.  Interestingly, we find that there
is a novel shape for vector RW solution. The density distribution
shapes of the localized waves in $\psi_1$ and $\psi_3$ are quite
different from the well-known eye-shaped one. There are two humps
and two valleys around a center, and the center's value is almost
equal to that of the background, as shown in Fig. 1(a) and (c). This
structure can be called as four-petaled structure in
temporal-spatial distribution. Moreover, the humps or valleys in
$\psi_1$ correspond to the valleys or humps in $\psi_3$. However,
the density distribution in $\psi_2$ is similar to the eye shaped RW
in single-component system, as shown in Fig. 1(b).  Therefore, the
whole density is still the well known eyes shape, as shown in Fig.
1(d). The novel shape should come from the cross phase modulation
effects, since the shape can not be observed in scalar RW. For
two-component coupled systems, it has been found numerically that
there are dark RWs in one component of the coupled system in
\cite{Bludov}. The dark RW has been given exactly in our previous
paper in \cite{zhao2}. Based on these results, we expect that there
should be some novel structures in more than three modes coupled
systems.

\begin{figure*}[htb]
\centering
\subfigure[]{\includegraphics[height=45mm,width=50mm]{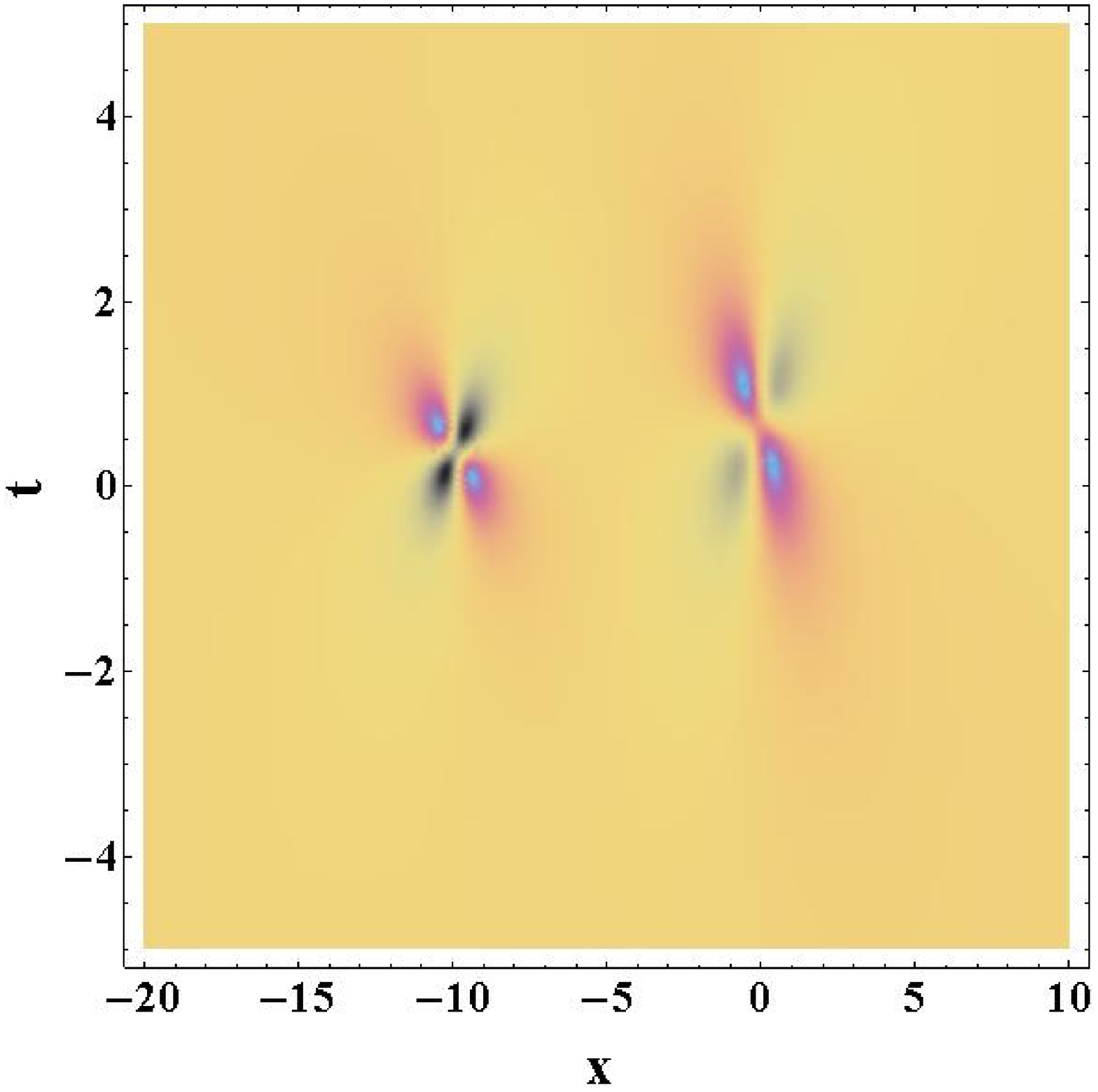}}
\hfil
\subfigure[]{\includegraphics[height=46mm,width=55mm]{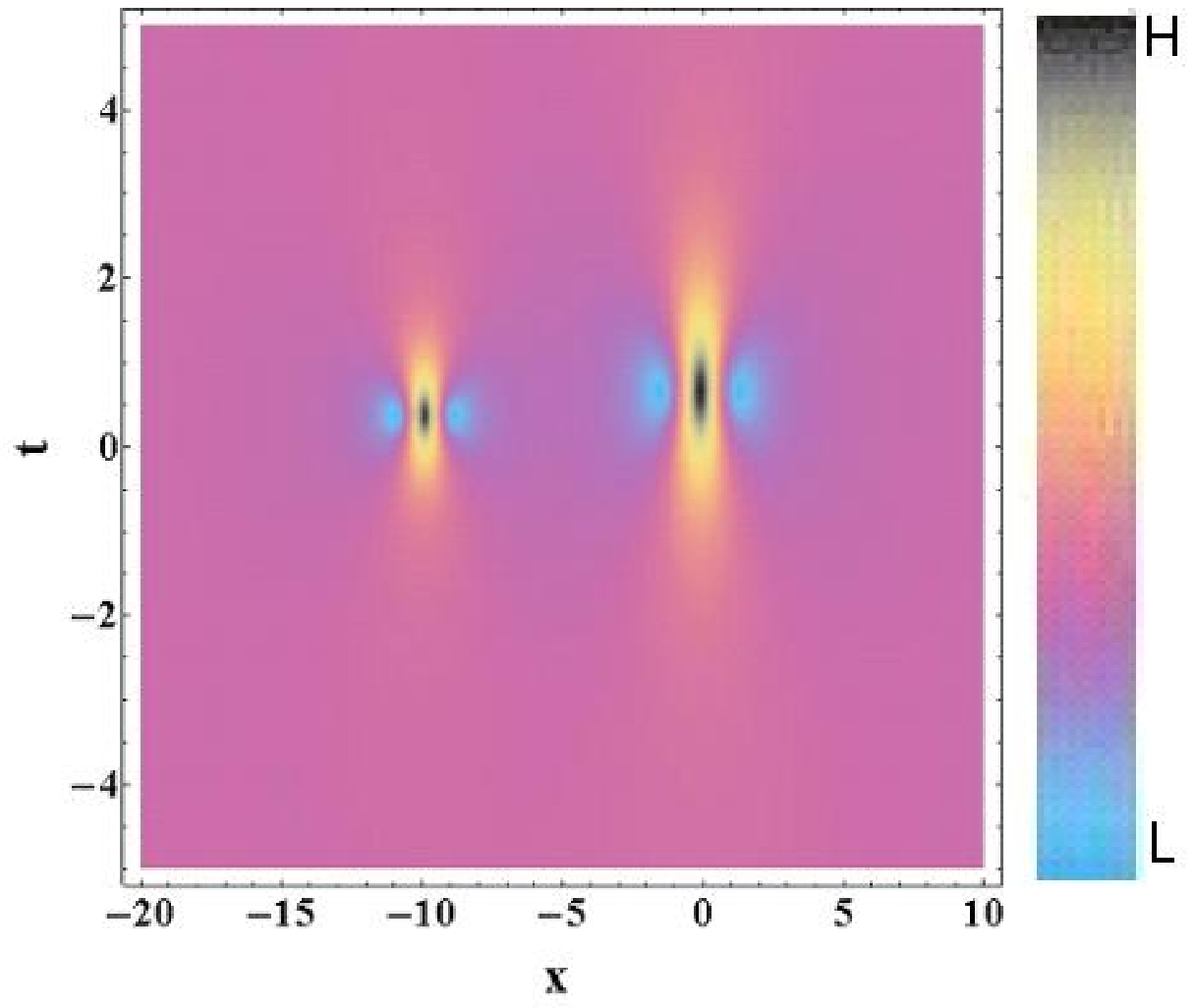}}
\hfil
\subfigure[]{\includegraphics[height=45mm,width=50mm]{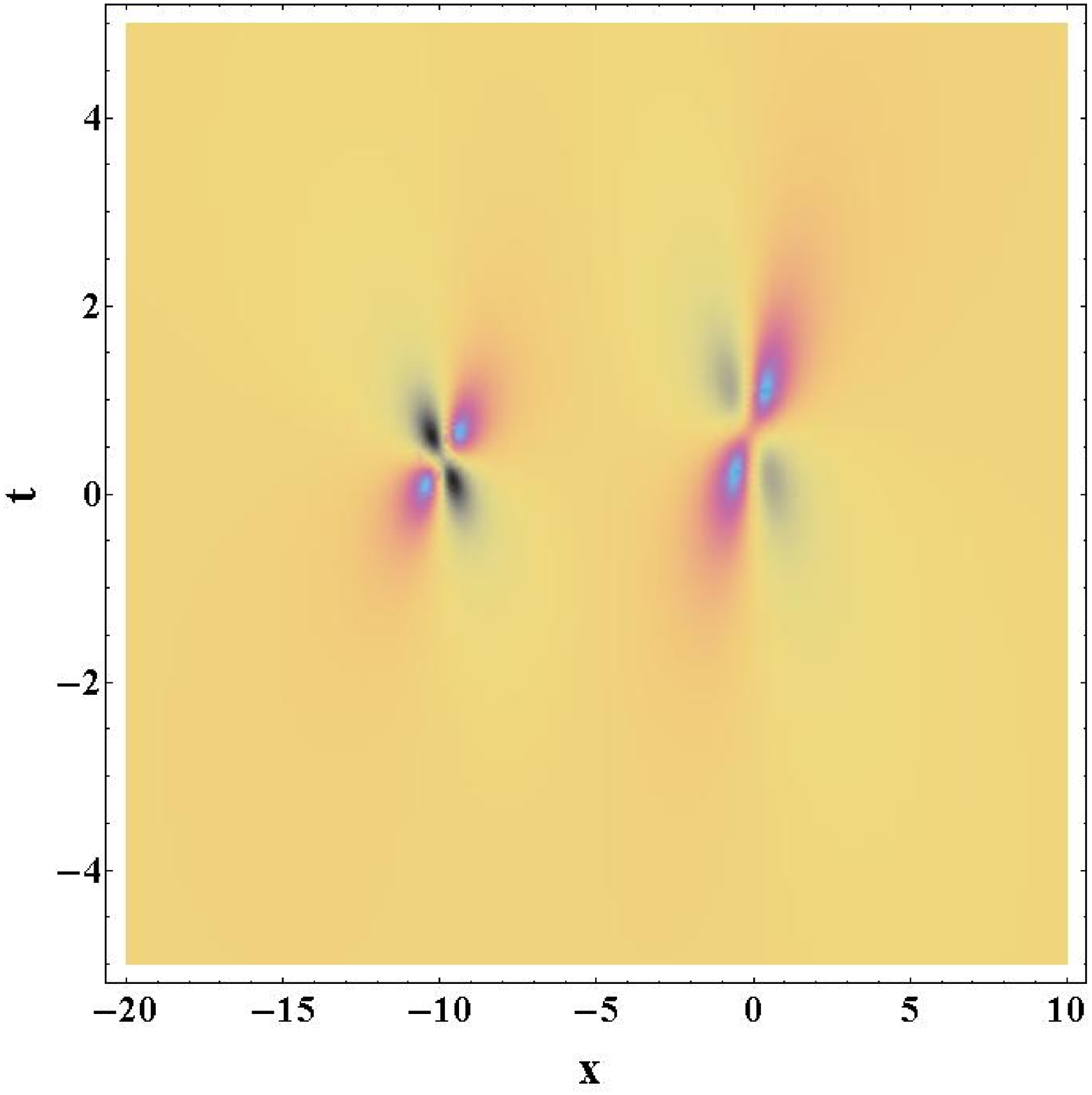}}
\hfil
\subfigure[]{\includegraphics[height=45mm,width=50mm]{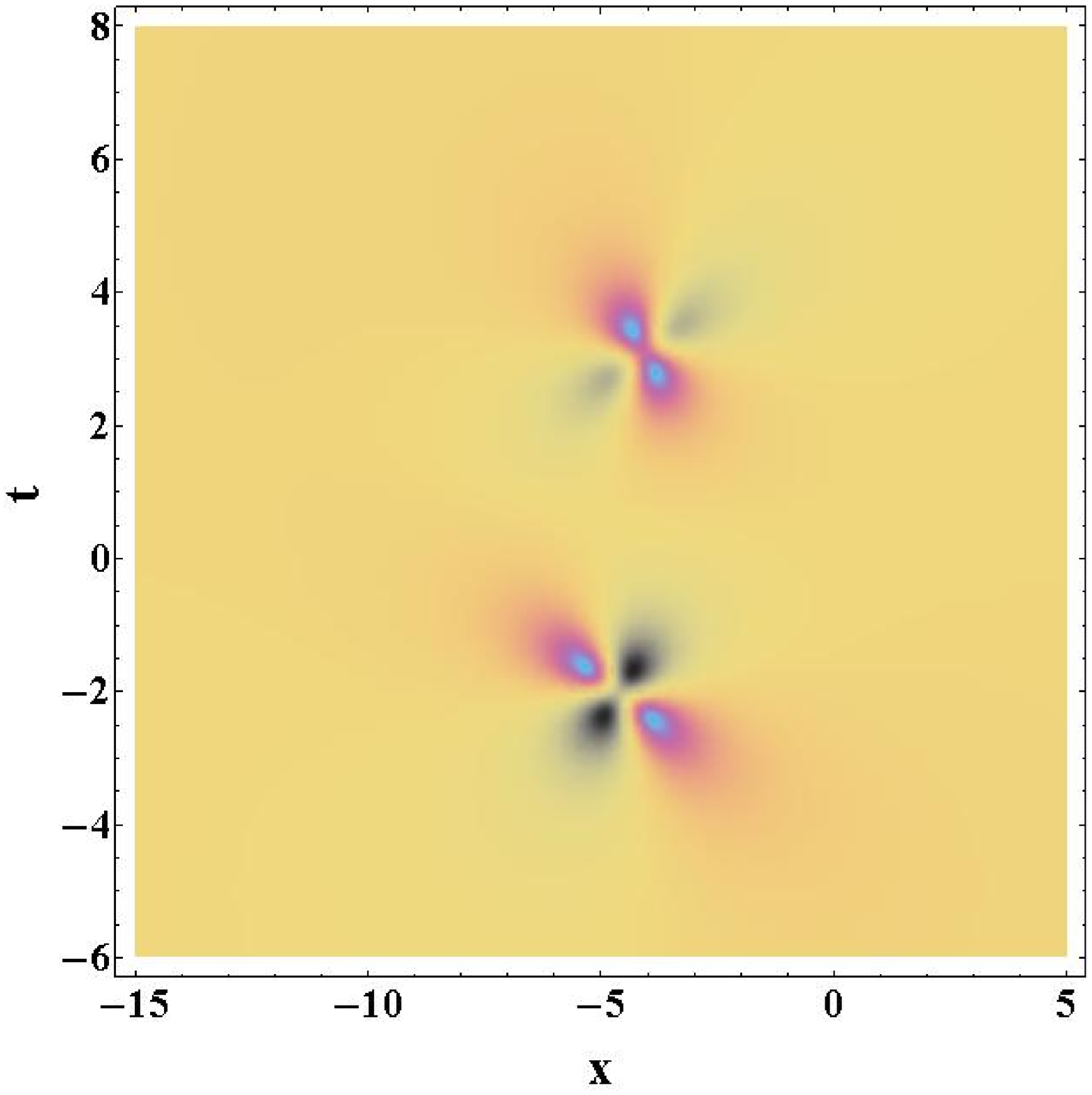}}
\hfil
\subfigure[]{\includegraphics[height=46mm,width=55mm]{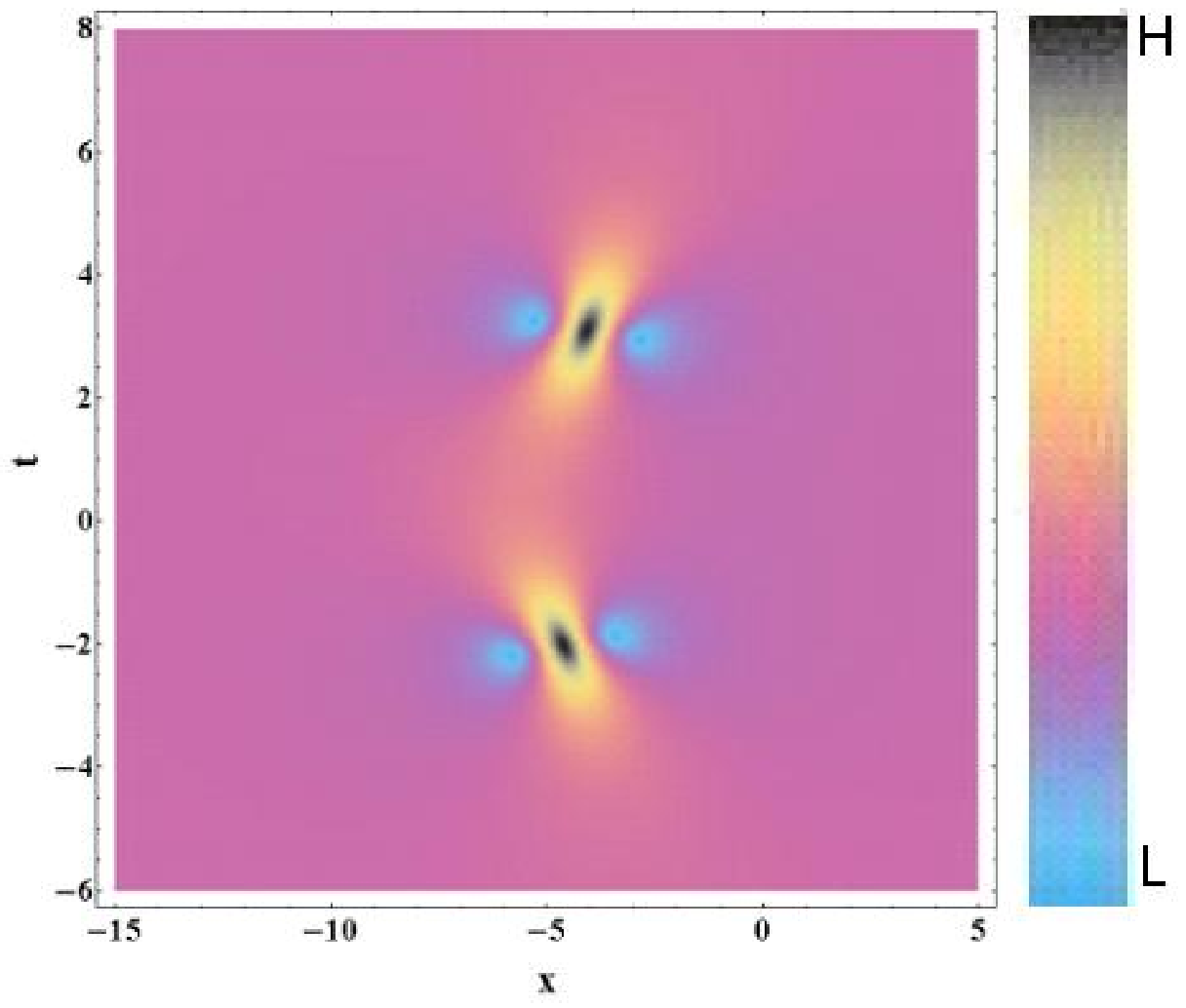}}
\hfil
\subfigure[]{\includegraphics[height=45mm,width=50mm]{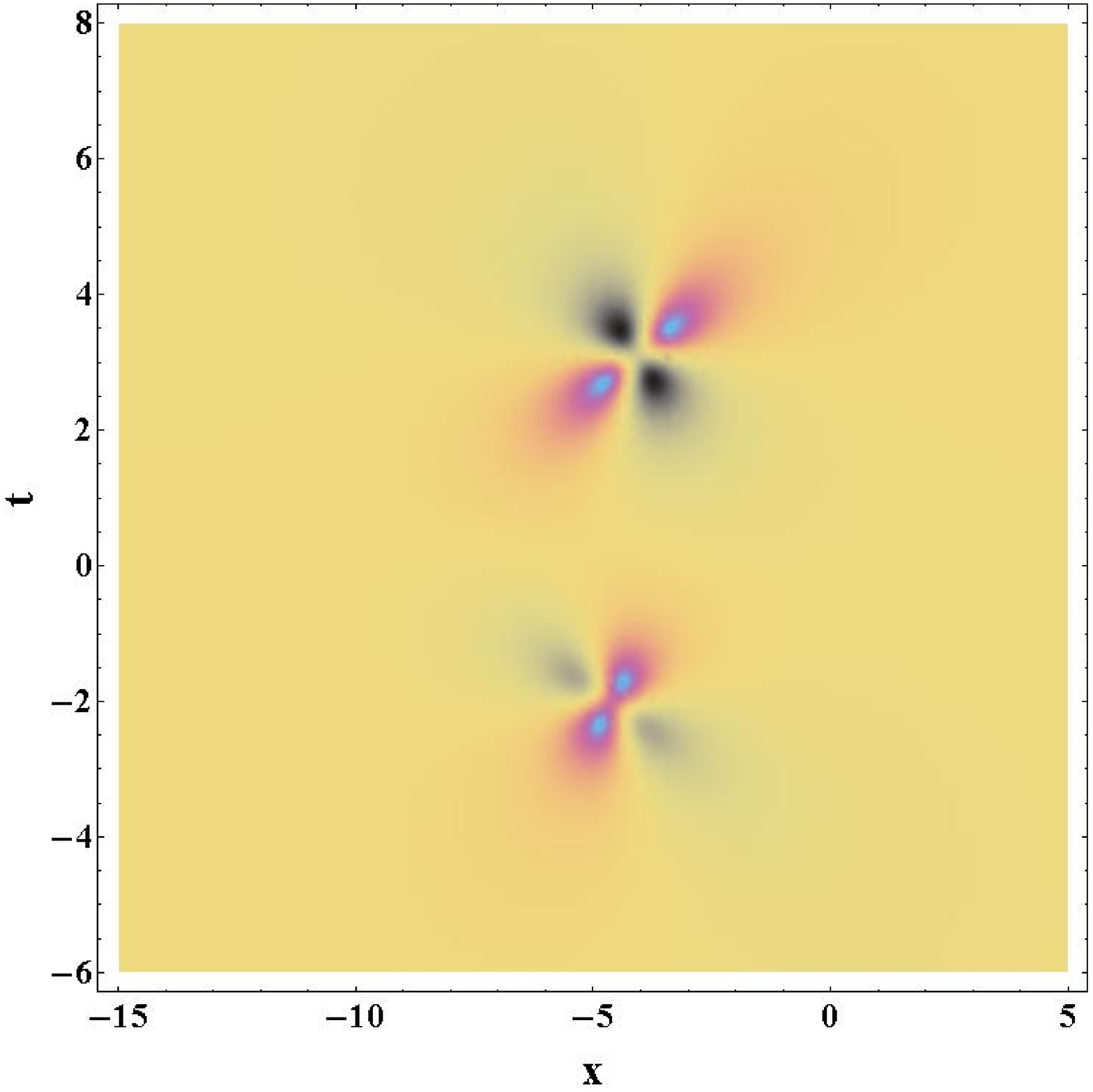}}
\caption{(color online)  The evolution plot of two vector RWs in
coupled system, (a) for the RWs in $\psi_1$ component, (b) for the
RWs in $\psi_2$ component. (c)  for the RWs in $\psi_3$ component.
The coefficients are $A_1=0, A_2=40, A_3=8$, and $A_4 = 0$. (d), (e)
and (f) show the evolution of RWs in $\psi_1$, $\psi_2$, and
$\psi_3$ respectively. The coefficients are $A_1=150, A_2=40,
A_3=8$, and $A_4 = 0$. }
\end{figure*}

\emph{ Two vector rouge waves}--- When $A_4=0$, there are two vector
RWs appearing in the temporal-spatial distribution, shown in Fig. 2.
When $A_1\leq 0$, there are two vector RWs emerging at a certain
propagation distance, as shown in Fig. 2(a)-(c). It is seen that the
structures of the two RWs in every mode are similar, and only the
sizes are different. The four-petaled structure RWs emerge in
$\psi_{1,3}$ and the eye-shaped ones emerge in $\psi_2$ component.
When $A_1>>0$, the two distinct RWs emerge at different propagation
distances, shown in Fig. 2(d)-(f). There is a rotation on $x-t$
distribution plane for the two RWs. Varying the parameter $A_{1,3}$,
one can observe the interactions between the two RWs.
\begin{figure*}[htb]
\centering
\subfigure[]{\includegraphics[height=45mm,width=50mm]{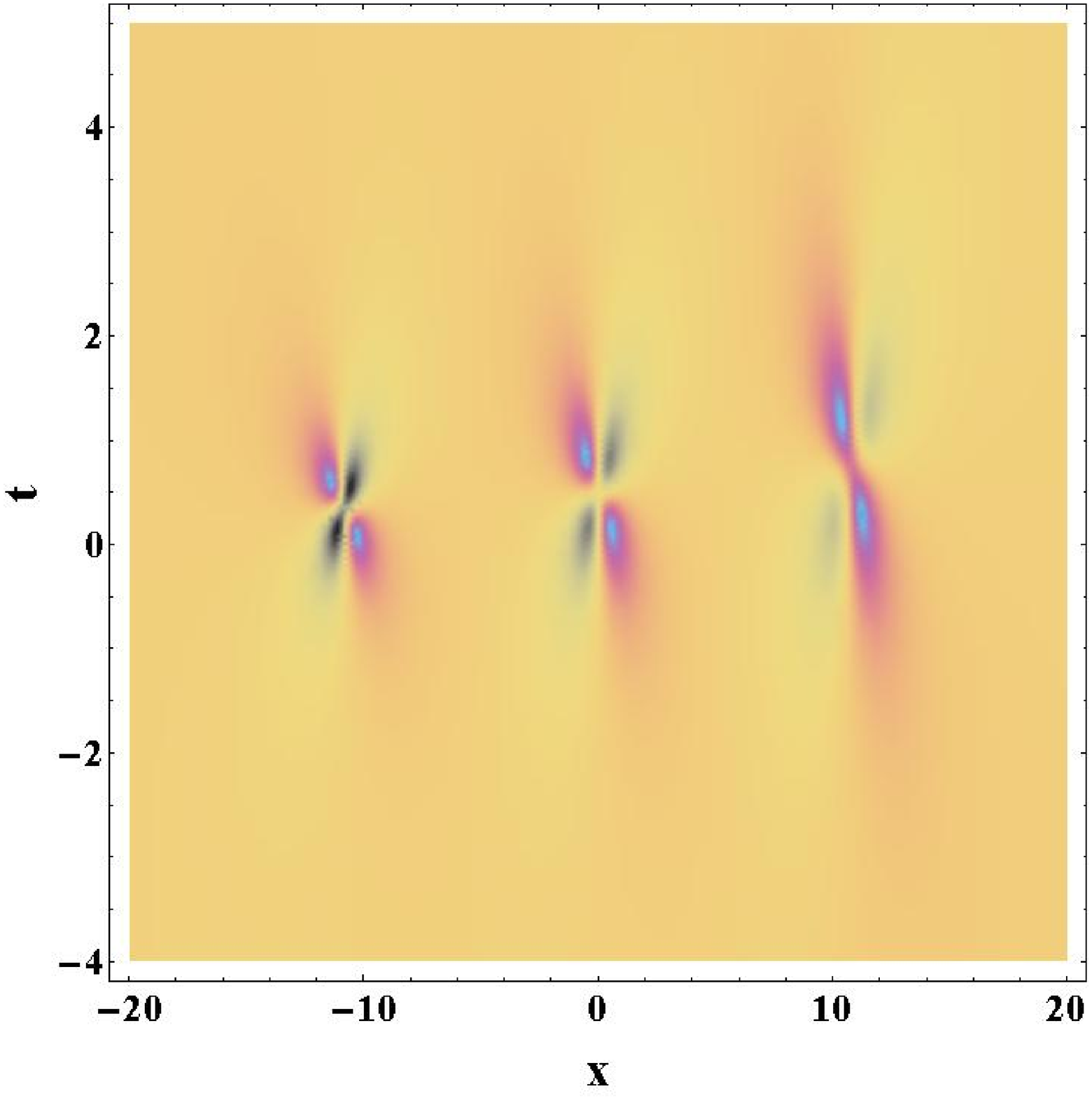}}
\hfil
\subfigure[]{\includegraphics[height=46mm,width=55mm]{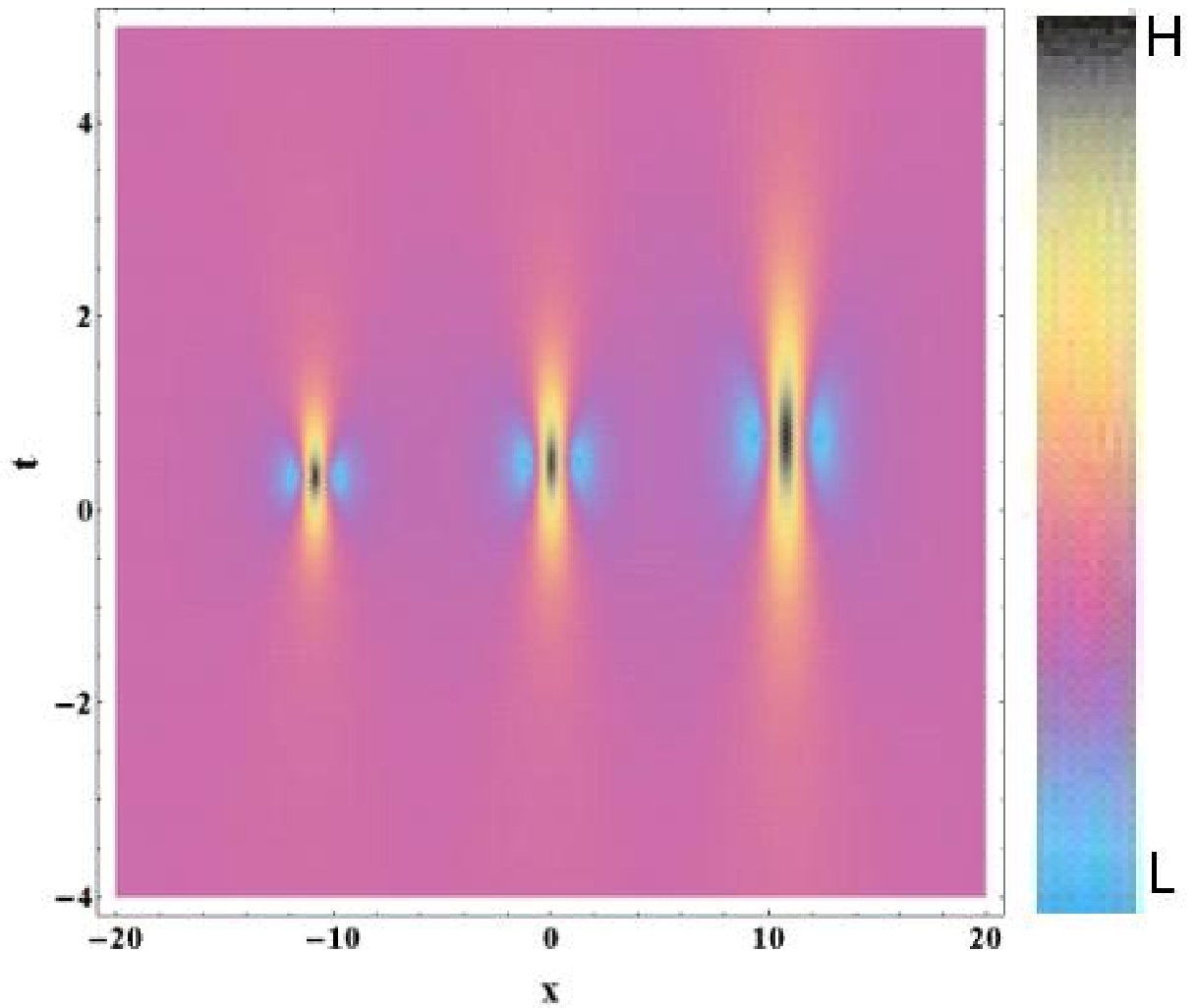}}
\hfil
\subfigure[]{\includegraphics[height=45mm,width=50mm]{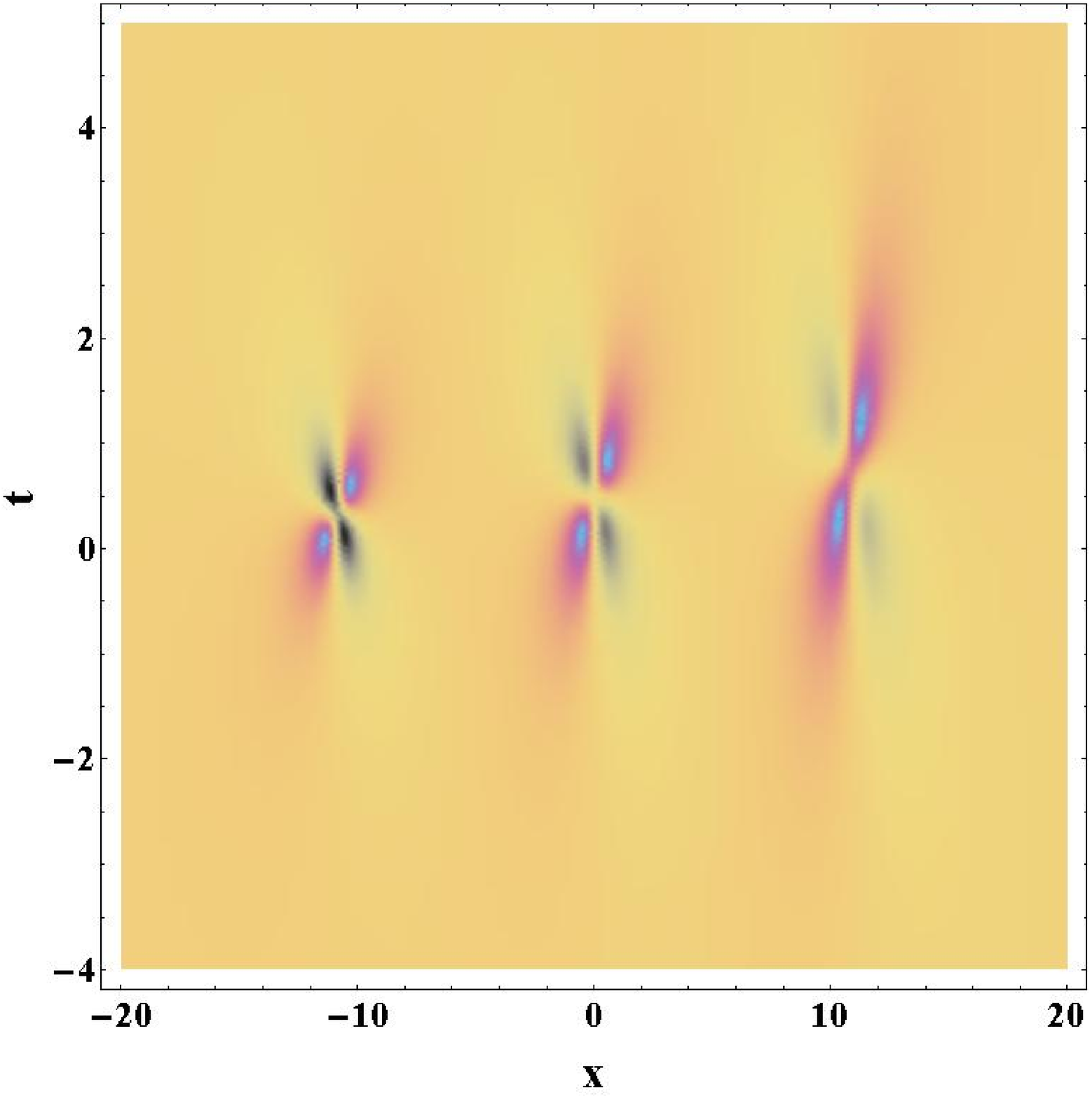}}
\hfil
\subfigure[]{\includegraphics[height=45mm,width=50mm]{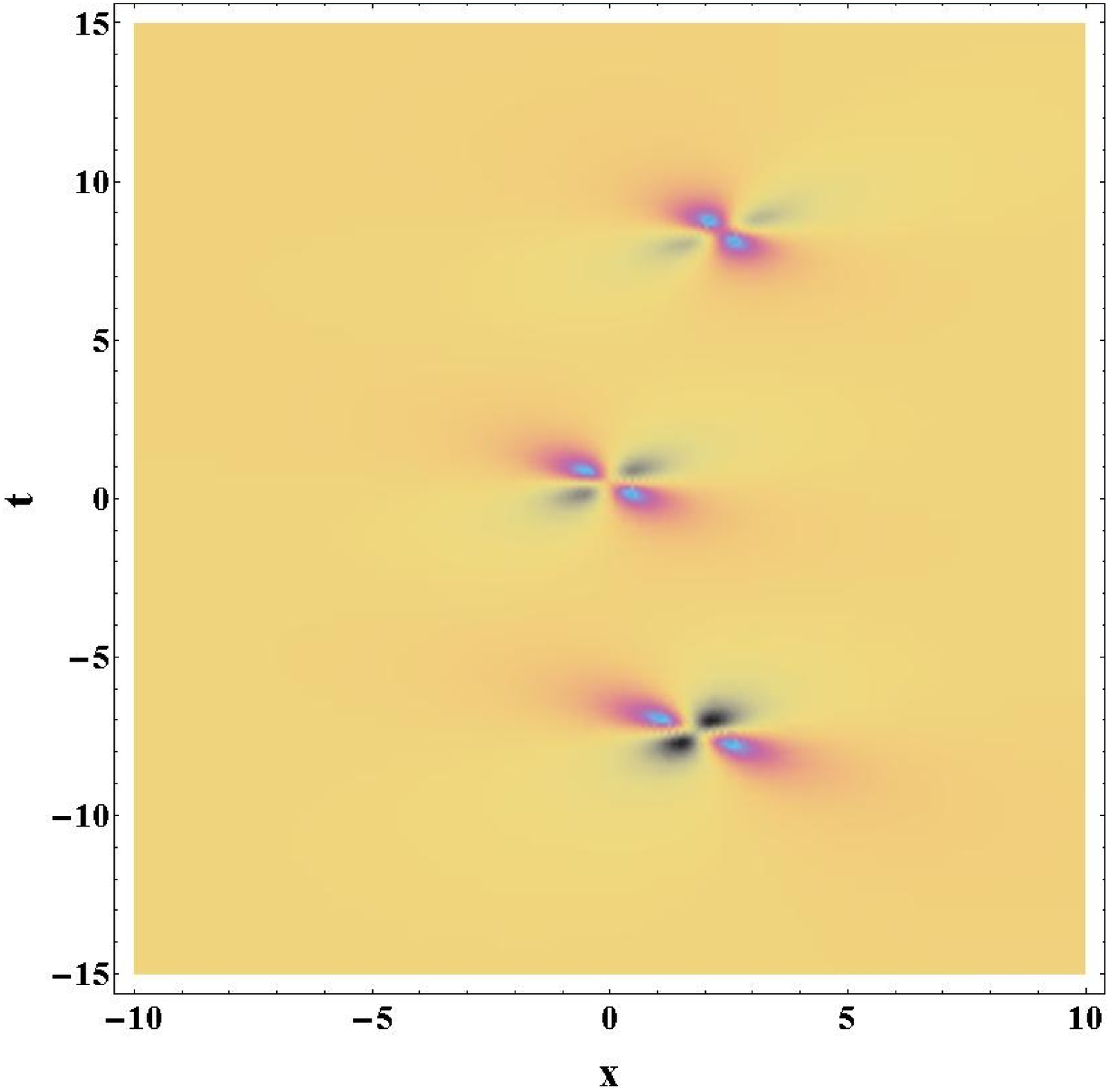}}
\hfil
\subfigure[]{\includegraphics[height=46mm,width=55mm]{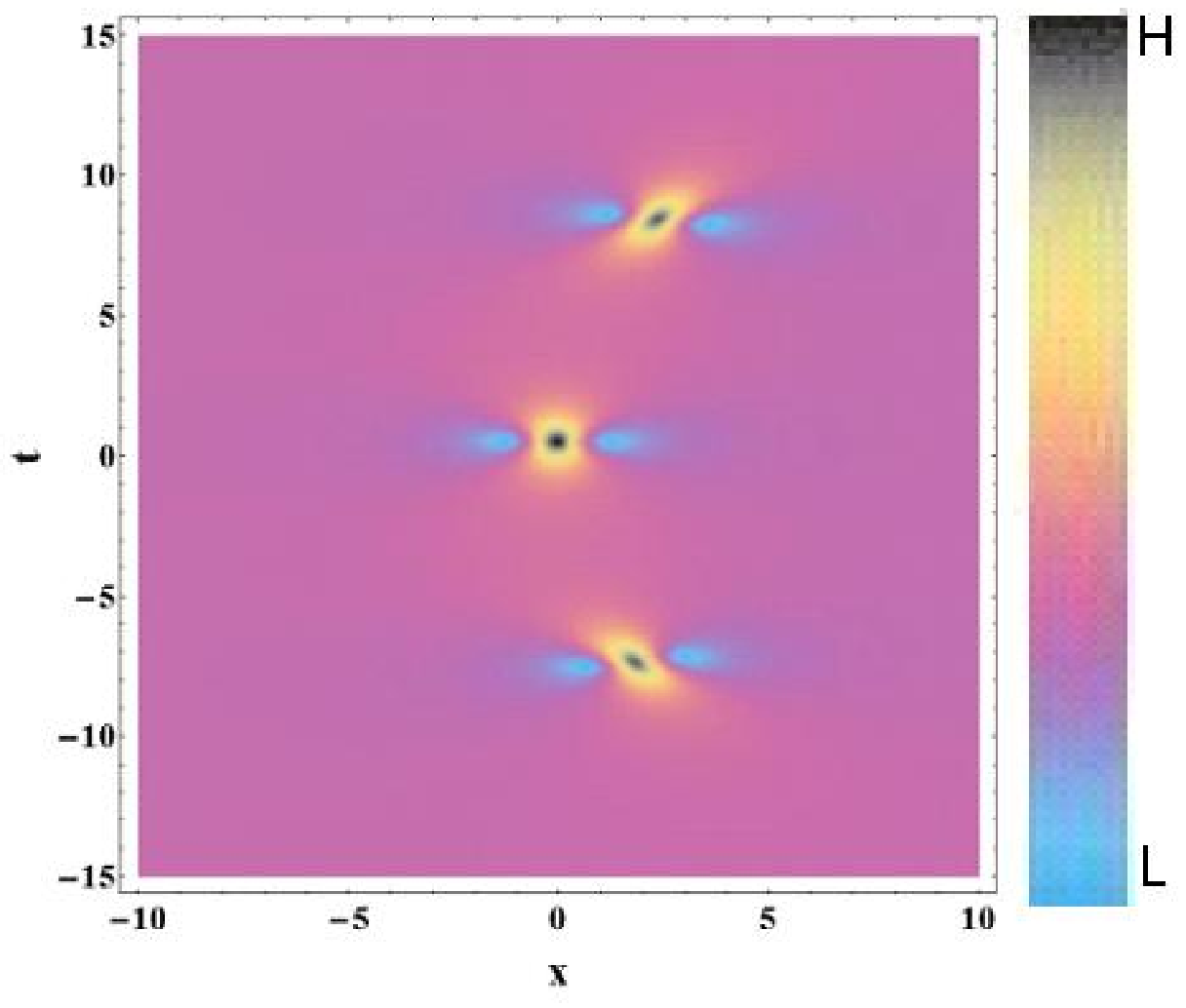}}
\hfil
\subfigure[]{\includegraphics[height=45mm,width=50mm]{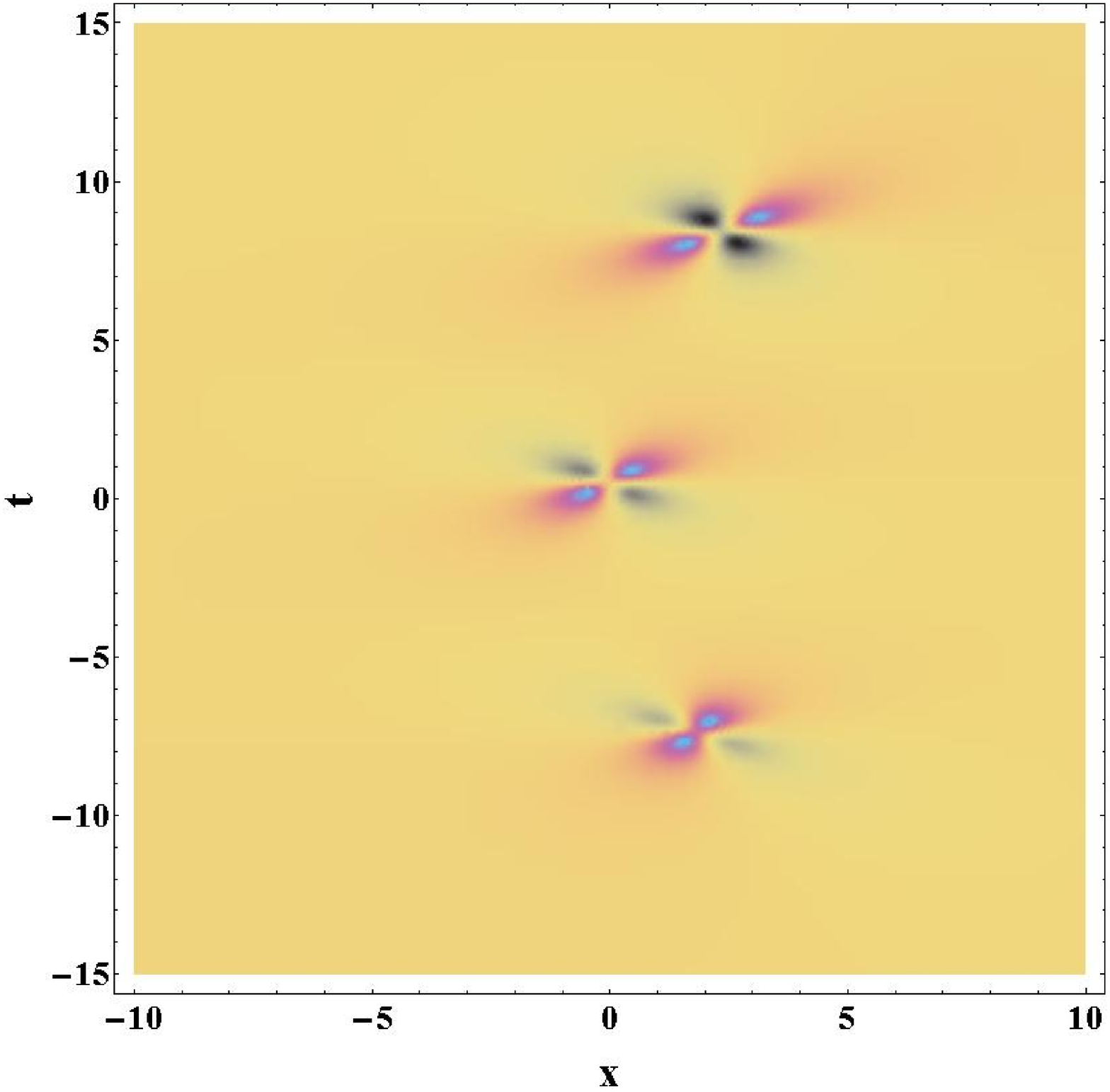}}
\caption{(color online) The evolution plot of three vector RWs in
coupled system, (a) for the RWs in $\psi_1$ component, (b) for the
RWs in $\psi_2$ component. (c)  for the RWs in $\psi_3$ component.
The coefficients are $A_1=0, A_2=20, A_3=0$, and $A_4 = -1$. (d),
(e) and (f) show the evolution of RWs in $\psi_1$, $\psi_2$, and
$\psi_3$ respectively. The coefficients are $A_1=0, A_2=20, A_3=0$,
and $A_4 =1$.}
\end{figure*}

\begin{figure}[htb]
\centering
\subfigure[]{\includegraphics[height=35mm,width=40mm]{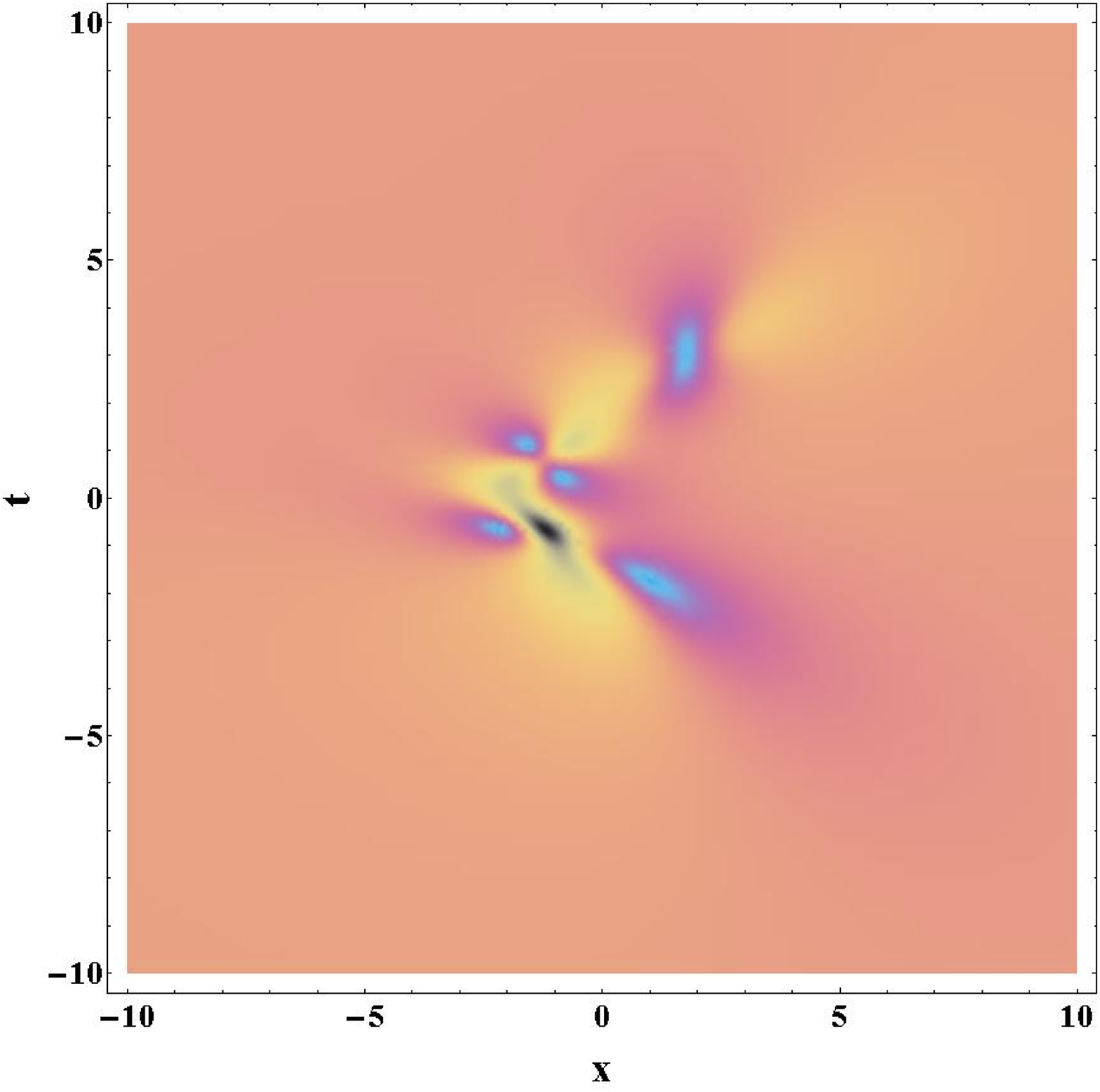}}
\hfil
\subfigure[]{\includegraphics[height=36mm,width=45mm]{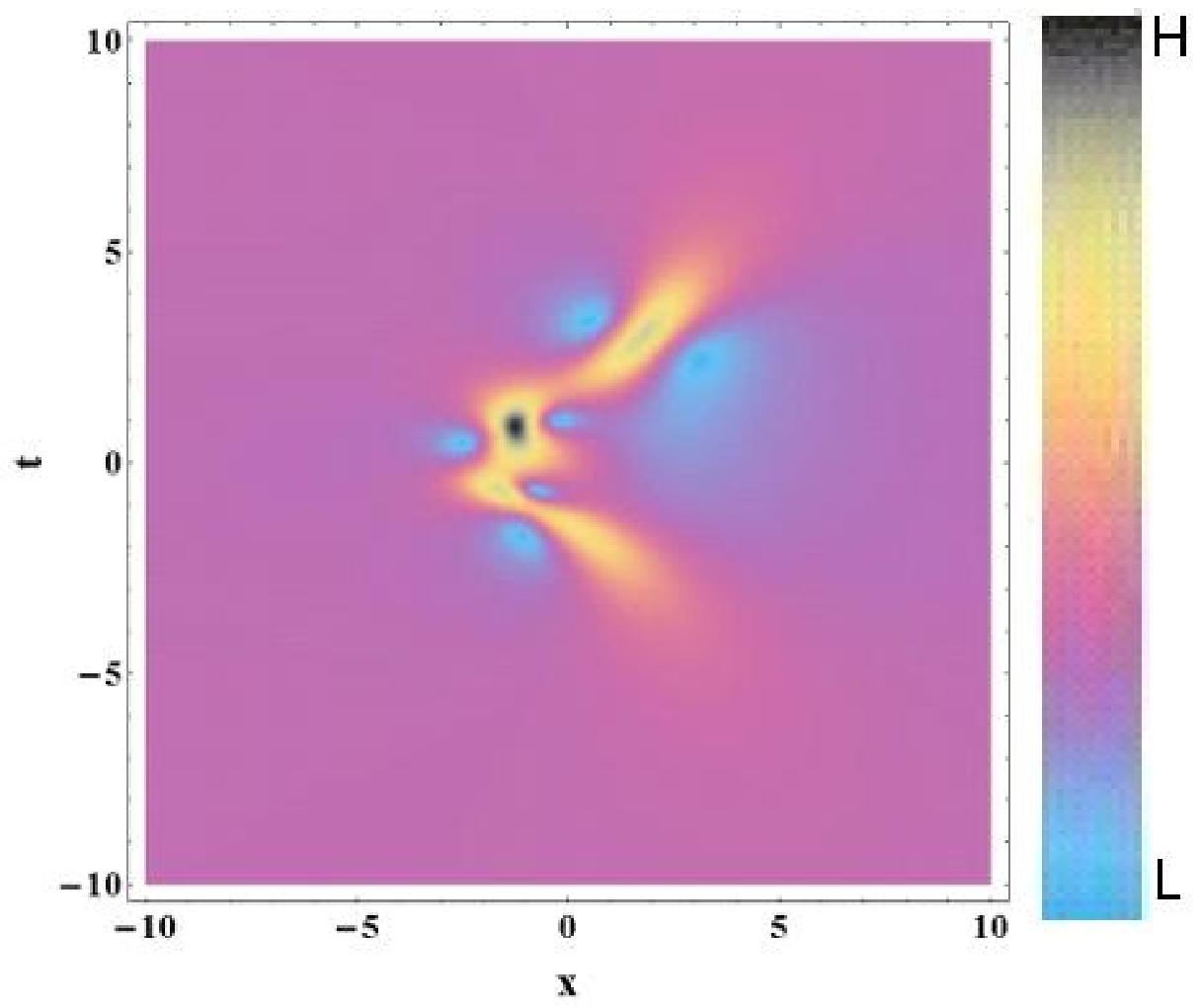}}
\hfil
\subfigure[]{\includegraphics[height=35mm,width=40mm]{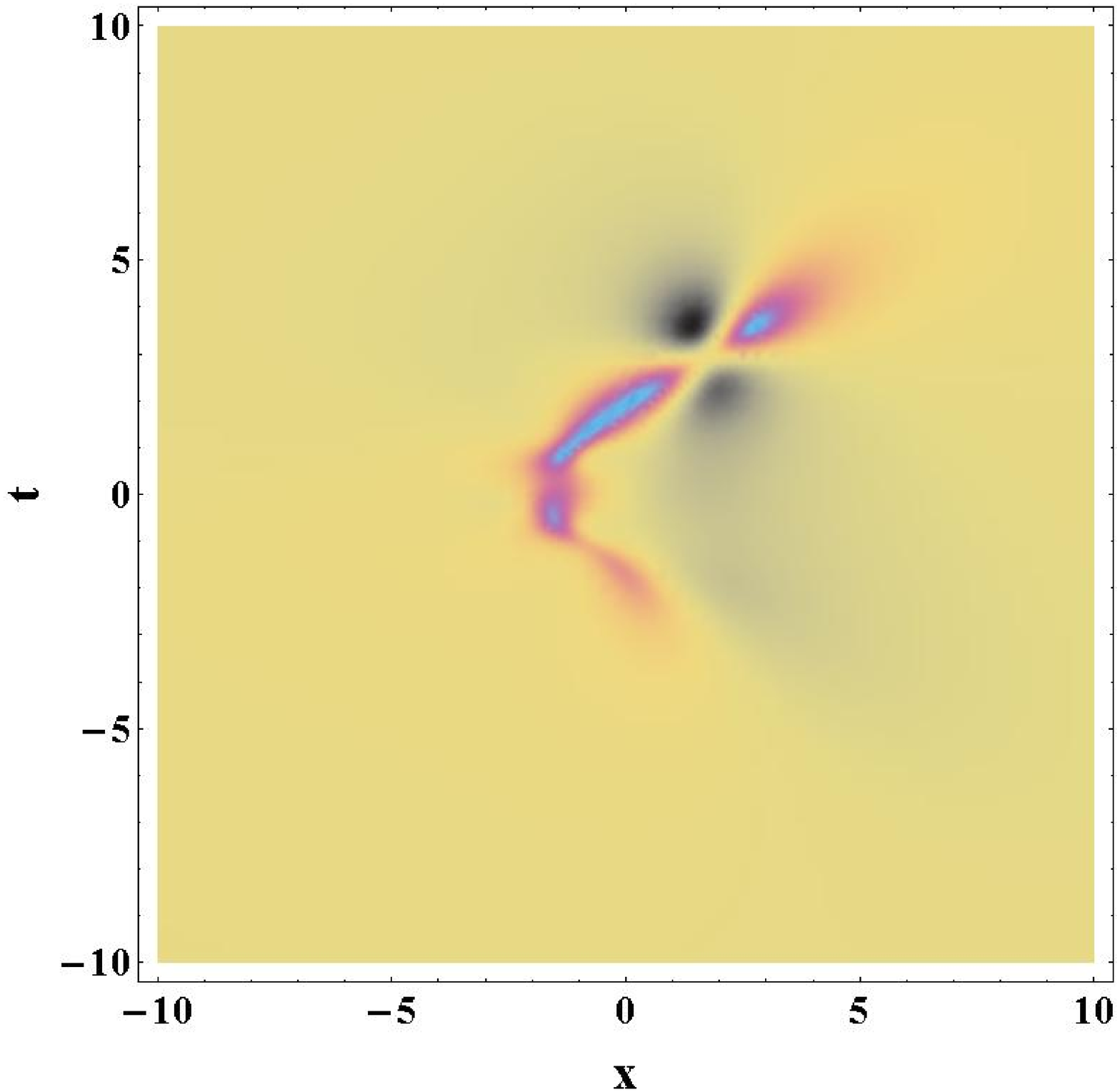}}
\hfil
\subfigure[]{\includegraphics[height=36mm,width=45mm]{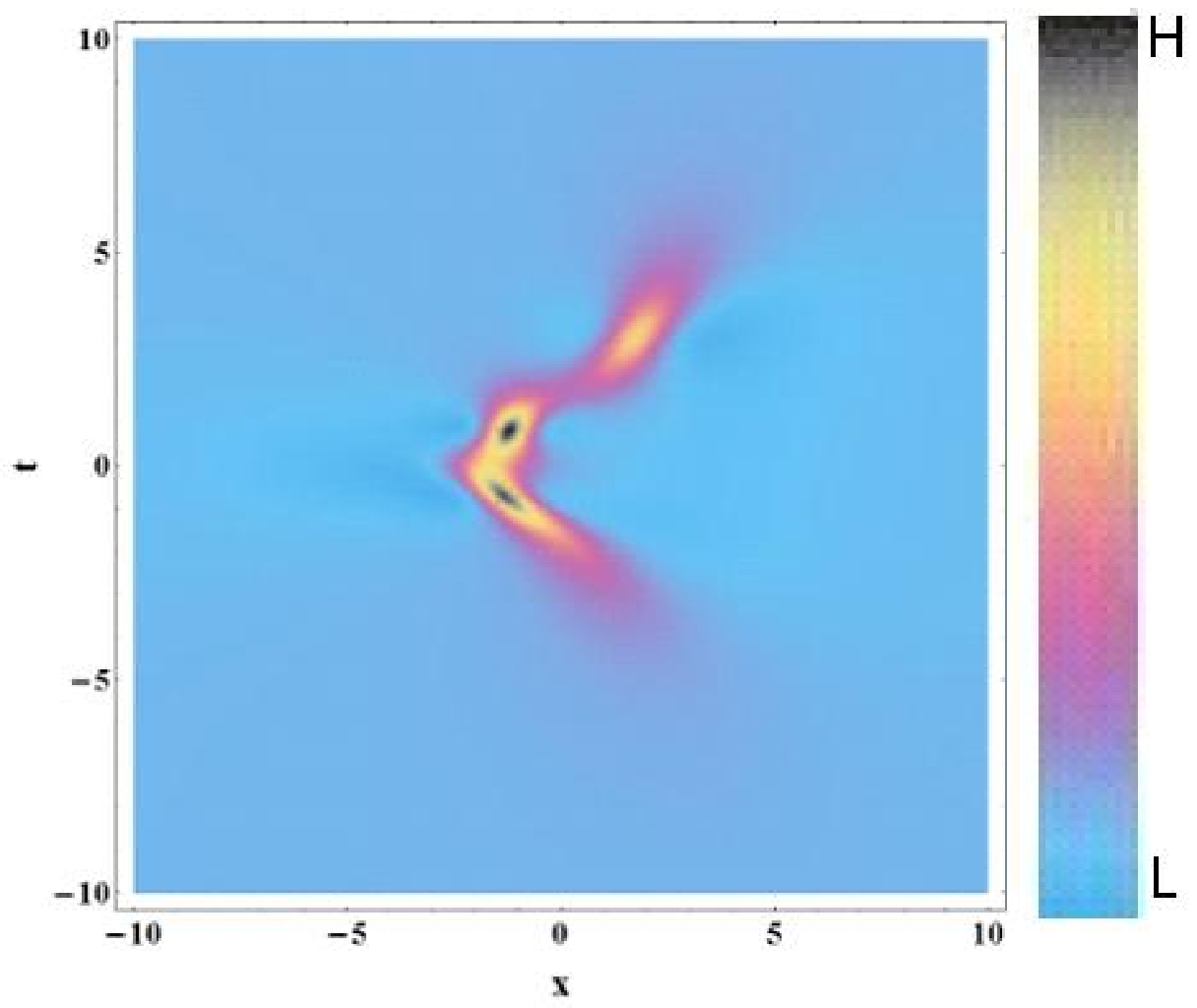}}
\caption{(color online)  The interaction plot of three RWs in the
coupled system, (a) for the RWs in $\psi_1$ component, (b) for the
RWs in $\psi_2$ component, (c) for the RWs in $\psi_3$ component.
(d) for the whole density of the three components. The coefficients
are $A_1 = 0, A_2 = 1, A_3 = 1$, and $ A_4 = 1$.}
\end{figure}

\emph{ Three vector rouge waves}--- When $A_4\neq0$, there are three
vector RWs appearing in the temporal-spatial distribution, as shown
in Fig. 3. When $A_4 A_2<0$ and $|A_2|>>|A_4|$, there are three
vector RWs emerging very clearly at a certain propagation distance,
as shown in Fig. 3(a)-(c). The structures of them in each component
are similar with different sizes. The valleys in $\psi_1$ still
correspond to the humps in $\psi_3$ component. There are three eye-
shaped RWs in $\psi_2$. When $A_4 A_2>0$ and $|A_2|>>|A_4|$, the
three RW emerge clearly at different propagation distances, as shown
in Fig. 3(d)-(f). The RW seems to have changing velocity on retarded
time. Varying the parameters, we can investigate the interaction
between these vector RWs conveniently. For example, making $A_2$
approach $A_4$, we can observe the interaction of the three RWs,
such as Fig. 4. It shows that RW's shape can be changed greatly
through interacting with others.  Two RWs almost fuse to one valley
in $\psi_3$ component with the condition, as shown in Fig. 4(c).

It should be noted that the distribution shapes of the two and three
RWs in the whole temporal-spatial distribution are very distinct
from the high-order RW in one-component system presented in
\cite{Ling,Akhmediev,Yang}. In the one-component systems, it is not
possible to observe just two RWs appearing in the whole
temporal-spatial distribution even for high-order RWs. Three RWs can
emerge on temporal-spatial distribution for second-order scalar RW,
but their distribution shapes are different from the three RWs
obtained here. Therefore, we call the generalized vector RW solution
as multi-vector RW.

\section{ Possibilities to observe these vector nonlinear waves}
Considering the experiments on RW in nonlinear fibers \cite{D.R.
Solli}, which have shown that the simple scalar NLS could describe
nonlinear waves in nonlinear fibers well,  we expect that these
different vector RWs could be observed in three-mode nonlinear
fibers.  One could introduce three-mode optical signals into a
nonlinear fiber operating in the anomalous group velocity dispersion
regime, marked by $\psi_j$ ($j=1,2,3$). Firstly, we assume the
nonlinear coefficients for the three modes are equal. The dispersion
and nonlinear coefficients are $1$ and $2$ in normalized units
respectively. Then, the backgrounds can be given by Eq.(5) and (6).
Explicitly, the amplitude of $\psi_1$ is set to be $s_1$, the
amplitude of $\psi_2$ is $s_2=\frac{\sqrt{2}}{2} s_1$, and the
amplitude of $\psi_3$ is $s_3=s_1$. The wave vector of $\psi_2$
marked by $k_2$ is set to be zero, and the wave vectors of $\psi_1$
and $\psi_3$ are $k_1=-k_3$, and $k_3=\frac{\sqrt{2}}{2} s_1$. Then
the initial optical signals are given by the presented vector RW
solution Eq.(7)-(9). For nonlinear fibers, the coordinates $x$ and
$t$ should be changed to be $t$ and $z$, which are retarded time and
propagation distance respectively. Under the corresponding
conditions, one, two, or three vector RWs could be observed in the
nonlinear fiber.

On the other hand, it is well known that there are spatial optical
solitons in planar waveguide. The similar conditions can be derived
directly through coordinates transformation for the coupled NLS in a
planar waveguide \cite{Cambournac}. The RWs in multi-mode planar
waveguide could be observed too. Additionally, considering the
studies on multi-components Bose-Einstein condensates
\cite{Das,Becker,Engels}, we expect that the vector RWs could also
be observed with the condition.

\section{Discussion and Conclusion}
  In summary, we find some novel structures for vector RWs in a three-component
coupled system. The novel shape lies in that there are two humps and
two valleys around one center in the temporal-spatial distribution.
The obtained multi-vector RW solution can be used to describe one,
two, and even three vector RWs in the coupled system, whose density
distribution shapes are different from high-order scalar RW. The
corresponding conditions for their emergence are presented
explicitly.  Under these conditions, the ideal initial signals for
them can be given from the generalized rational solution, which
could be helpful for experimental observation. Based on these
results, we expect that abundant novel structures could exist in
more-than-three-component coupled systems.

The coupled system can be used to describe three-component BEC,
multi-mode optical transmission, and so on. Therefore, we believe
that these nonlinear waves can be observed experimentally. As an
example, the possible way to observe vector RWs in three-mode
nonlinear fiber is discussed here. Recently,
 the high-order modulation instability with RWs has been observed
 in nonlinear fiber optics in \cite{Erkintalo}. There are many
 possibilities to observe similar phenomena in coupled nonlinear
 fiber system.

\section*{Acknowledgments}
 This work is supported by the National Fundamental Research Program of China
(Contact No. 2011CB921503), the National Science Foundation of China
(Contact Nos. 11274051, 91021021).

\section*{Appendix A: The analytic expressions for $H_j(x,t)$ and $G_j(x,t)$}
\begin{figure*}
\begin{eqnarray*}
H_1&=&-(1+i) \left[6 i A_1+(6-6 i) A_3-6 i \sqrt{2}A_4+(6+12 i) A_3
t+6 \sqrt{2} A_4 t-6 i A_3 t^2-(6-6 i)
\sqrt{2} A_4 t^2+2 \sqrt{2}A_4 t^3\right.\\
&&\left.+A_2 (6 \sqrt{2}-6 \sqrt{2} t+6 i x)+6 \sqrt{2} A_3 x+(6-6
i)A_4 x-6 \sqrt{2} A_3 t x+(6+12 i) A_4 t x-6 i A_4 t^2 x+3 i A_3
x^2\right.\\
&&\left.+3 \sqrt{2} A_4x^2-3 \sqrt{2}A_4 t x^2+i A_4 x^3\right]
\left[6 \sqrt{2} A_1+(6+12 i) \sqrt{2}A_3 t-(6+6 i) A_4 t-6 \sqrt{2}
A_3 t^2+(18-6 i)
A_4 t^2\right.\\
&&\left.+4 i A_4 t^3-(6-6 i) A_3 x-12 i A_3 t x+(6+12 i) \sqrt{2}
A_4 t x-6 \sqrt{2} A_4 t^2 x+3 \sqrt{2} A_3 x^2-(3-3 i) A_4 x^2-6 i
A_4 t x^2\right.\\
&&\left.+\sqrt{2}A_4 x^3+6A_2 (-1+i-2 i t+\sqrt{2} x)\right],\nonumber\\
G_1&=&\sqrt{2} \left[36A_1^2+36 A_3^2+54 \sqrt{2} A_3 A_4+54
A_4^2+108 A_3^2 t^2-36 \sqrt{2} A_3 A_4 t^2-72 A_3^2 t^3+48 \sqrt{2}
A_3 A_4 t^3-24A_4^2 t^3\right.\\
&&\left.+36 A_3^2 t^4-48 \sqrt{2}A_3 A_4 t^4+108 A_4^2 t^4-24A_4^2
t^5+8 A_4^2 t^6+36 \sqrt{2}A_3^2 x+108 A_3 A_4 x+54 \sqrt{2}A_4^2
x-72\sqrt{2}A_3^2 t^2 x\right.\\
&&\left.+144 A_3A_4 t^2 x-36 \sqrt{2}A_4^2 t^2 x-48A_3A_4 t^3 x+48
\sqrt{2}A_4^2 t^3 x+24A_3 A_4 t^4 x-48 \sqrt{2}A_4^2 t^4 x+36 A_3^2
x^2+54 \sqrt{2}A_3A_4 x^2\right.\\
&&\left.+54 A_4^2 x^2-36A_3^2 t x^2+36A_3^2 t^2 x^2-72
\sqrt{2}A_3A_4 t^2 x^2+72 A_4^2 t^2 x^2-24A_4^2 t^3 x^2+12 A_4^2 t^4
x^2+36 A_3 A_4 x^3\right.\\
&&\left.+18 \sqrt{2}A_4^2 x^3-24A_3A_4 t x^3+24A_3A_4 t^2 x^3-24
\sqrt{2} A_4^2 t^2 x^3+9 A_3^2 x^4+9A_4^2 x^4-6 A_4^2 t x^4+6A_4^2
t^2 x^4+6A_3 A_4 x^5\right.\\
&&\left.-12A_2 ( -3A_4+6A_4 t^2-8A_4 t^3+4A_4 t^4-3 A_4\sqrt{2} x-3
A_4x^2-A_4x^4+ 6 A_3\sqrt{2}t^2-6A_3 t^2 x+6A_3 t x-3\sqrt{2} A_3)\right.\\
&&\left.+12 A_1(6 A_2 x+ 6A_3 t-6A_3 t^2+3A_3 x^2+ 6\sqrt{2}
A_4t^2-6A_4
t^2x+6A_4 t x+A_4x^3 )\right.\\
&&\left.+72 A_2A_3 x+36 A_2 A_3x^3+A_4^2 x^6+36 A_2^2 (1-2 t+2
t^2+x^2) \right ].
\end{eqnarray*}
\end{figure*}

\begin{figure*}
\begin{eqnarray*}
 H_2&=&72 \sqrt{2}A_1^2+72 i
\sqrt{2}A_2^2+(72-72 i) \sqrt{2} A_1A_3+144 iA_2 A_3+(144-144 i)
A_1A_4+144 i \sqrt{2}A_2
A_4-(144+144 i) \sqrt{2} A_2^2 t\nonumber\\
&&+(144+144 i) \sqrt{2}A_1 A_3 t+(216+72 i) \sqrt{2} A_3^2
t-(288+144 i) \sqrt{2} A_2 A_4 t+(288+144 i)A_3 A_4 t-144
\sqrt{2}A_4^2 t+144 \sqrt{2} A_2^2 t^2\nonumber\\
&&-144 \sqrt{2}A_1 A_3 t^2-288 A_2 A_3 t^2+216 i \sqrt{2} A_3^2
t^2+288 A_1 A_4 t^2+(144-288 i) \sqrt{2} A_2 A_4 t^2-144 A_3A_4
t^2\nonumber\\
&&+(144-360 i) \sqrt{2} A_4^2 t^2-(144+144 i) \sqrt{2}A_3^2
t^3+(192+192 i) \sqrt{2} A_2 A_4 t^3+(192+192 i)A_3 A_4 t^3+(192+48
i) \sqrt{2}A_4^2 t^3\nonumber\\
&&+72 \sqrt{2}A_3^2 t^4-96 \sqrt{2}A_2 A_4 t^4-192 A_3A_4
t^4+(96+120 i) \sqrt{2}A_4^2 t^4-(48+48 i) \sqrt{2}A_4^2 t^5+16
\sqrt{2}A_4^2 t^6+144 \sqrt{2}A_1 A_2 x\nonumber\\
&&+(72+72 i) \sqrt{2} A_2A_3 x+144 i A_3^2 x+(72-72 i) \sqrt{2}A_1
A_4 x+144 A_2A_4 x+144 i \sqrt{2}A_3 A_4x-(144+144 i) \sqrt{2} A_2
A_3 t x\nonumber\\
&&+(144+144 i) \sqrt{2} A_1 A_4 t x+144 \sqrt{2} A_3A_4 t x+(288+144
i) A_4^2 t x+144 \sqrt{2}A_2 A_3 t^2 x-288 A_3^2 t^2 x\nonumber\\
&&-144 \sqrt{2}A_1A_4 t^2 x+(144+144 i) \sqrt{2}A_3A_4 t^2x-144A_4^2
t^2 x-(96+96 i) \sqrt{2} A_3A_4 t^3 x+(192+192 i) A_4^2
t^3 x\nonumber\\
&&+48 \sqrt{2}A_3 A_4 t^4 x-192A_4^2 t^4 x+72 \sqrt{2} A_2^2 x^2+72
\sqrt{2}A_1A_3 x^2+(36+36 i) \sqrt{2} A_3^2 x^2+72 \sqrt{2}A_2 A_4 x^2\nonumber\\
&&+(72+144 i)A_3 A_4 x^2+72 i \sqrt{2}A_4^2 x^2-(72+72 i) \sqrt{2}
A_3^2 t x^2+72 \sqrt{2}A_4^2 t x^2+72 \sqrt{2}A_3^2 t^2 x^2+12 \sqrt{2}A_4^2 t^2 x^4\nonumber\\
&&-288 A_3 A_4 t^2 x^2+(72+72 i) \sqrt{2} A_4^2 t^2 x^2-(48+48 i)
\sqrt{2}A_4^2 t^3 x^2+24 \sqrt{2}A_4^2 t^4 x^2+72 \sqrt{2} A_2A_3
x^3+2 \sqrt{2}A_4^2 x^6\nonumber\\
&&+24 \sqrt{2}A_1A_4 x^3+(48+24 i) \sqrt{2}A_3 A_4 x^3+(24+48 i)
A_4^2 x^3-(48+48 i) \sqrt{2}A_3 A_4 t x^3+48 \sqrt{2}A_3 A_4 t^2
x^3\nonumber\\
&&-96A_4^2 t^2 x^3+18 \sqrt{2}A_3^2 x^4+24 \sqrt{2} A_2 A_4
x^4+(12+6 i) \sqrt{2} A_4^2 x^4-(12+12 i) \sqrt{2}A_4^2 t x^4+12
\sqrt{2}A_3A_4 x^5,\nonumber\\
G_2&=&36 \sqrt{2} A_1^2+36 \sqrt{2} A_2^2+72 A_2 A_3+36 \sqrt{2}
A_3^2+36 \sqrt{2}A_2 A_4+108A_3A_4+54 \sqrt{2}A_4^2-72 \sqrt{2}
A_2^2 t\nonumber\\
&&+72 \sqrt{2}
 A_1 A_3t+72 \sqrt{2}A_2^2 t^2-72 \sqrt{2}
 A_1A_3 t^2-144 A_2 A_3 t^2+108 \sqrt{2}
A_3^2 t^2+12 \sqrt{2}A_2A_4 x^4\nonumber\\
&&+144  A_1A_4 t^2-72 \sqrt{2} A_2A_4t^2-72 A_3A_4 t^2-72
\sqrt{2}A_3^2 t^3+96 \sqrt{2} A_2A_4 t^3+96 A_3A_4 t^3-24
\sqrt{2}A_4^2 t^3\nonumber\\
&&+36 \sqrt{2}A_3^2 t^4-48 \sqrt{2}A_2A_4 t^4-96A_3 A_4 t^4+108
\sqrt{2}A_4^2 t^4-24 \sqrt{2} A_4^2 t^5+8 \sqrt{2}A_4^2 t^6+72
\sqrt{2} A_1 A_2 x\nonumber\\
&&+72 \sqrt{2}A_2 A_3 x+72A_3^2 x+72 A_2 A_4 x+108 \sqrt{2} A_3 A_4
x+108 A_4^2 x
-72 \sqrt{2} A_2A_3 t x+72 \sqrt{2} A_1A_4 t x\nonumber\\
&&+72 \sqrt{2}A_2A_3 t^2 x-144 A_3^2 t^2 x-72 \sqrt{2}  A_1 A_4 t^2
x +144 \sqrt{2} A_3A_4 t^2 x-72A_4^2 t^2 x-48 \sqrt{2}A_3 A_4 t^3
x\nonumber\\
&&+96 A_4^2 t^3 x+24 \sqrt{2}A_3A_4 t^4 x-96 A_4^2 t^4 x+36
\sqrt{2}A_2^2 x^2+36 \sqrt{2}A_1 A_3 x^2+36 \sqrt{2}A_3^2 x^2+\sqrt{2}A_4^2 x^6\nonumber\\
&&+36 \sqrt{2}A_2A_4 x^2+108 A_3A_4 x^2+54 \sqrt{2} A_4^2 x^2-36
\sqrt{2}A_3^2 t x^2+36 \sqrt{2} A_3^2 t^2 x^2-144 A_3A_4 t^2 x^2\nonumber\\
&&+72 \sqrt{2}A_4^2 t^2 x^2-24 \sqrt{2}A_4^2 t^3 x^2+12 \sqrt{2}
A_4^2 t^4 x^2+36 \sqrt{2}A_2A_3 x^3+12 \sqrt{2}A_1A_4 x^3\nonumber\\
&&+36 \sqrt{2}A_3 A_4 x^3+36 A_4^2 x^3-24 \sqrt{2} A_3A_4 t x^3+24
\sqrt{2} A_3A_4 t^2 x^3-48 A_4^2 t^2 x^3+9 \sqrt{2}A_3^2 x^4\nonumber\\
&&+9 \sqrt{2} A_4^2 x^4-6 \sqrt{2} A_4^2 t x^4+6 \sqrt{2} A_4^2 t^2
x^4+6 \sqrt{2} A_3 A_4 x^5.
\end{eqnarray*}
\end{figure*}

\begin{figure*}
\begin{eqnarray*}
H_3&=&(1 + i) \left[6 A_1-6 iA_3 t-6 A_3 t^2+6 \sqrt{2}A_4 t^2-2 i
\sqrt{2} A_4 t^3+6 i \sqrt{2}A_3 t x-6 i A_4 t x-6 A_4 t^2 x+3 A_3
x^2+3 i \sqrt{2}A_4 t x^2\right.\\
&&\left.+A_4 x^3+6 A_2 \left(i \sqrt{2}t+x\right)\right] \left[6
\sqrt{2}A_1+(6+12 i) \sqrt{2}A_3 t-(6+6 i) A_4 t-6 \sqrt{2} A_3
t^2+(18-6 i) A_4 t^2+4 i A_4 t^3\right.\\
&&\left.-(6-6 i) A_3 x-12 i A_3 t x+(6+12 i) \sqrt{2} A_4 t x-6
\sqrt{2}A_4 t^2 x+3 \sqrt{2}A_3 x^2-(3-3 i)A_4 x^2\right.\\
&&\left.-6 i A_4 t x^2+\sqrt{2} A_4 x^3+6 A_2
\left((-1+i)-2 i t+\sqrt{2} x\right)\right],\nonumber\\
G_3&=&\sqrt{2} \left[36 A_1^2+36 A_3^2 +54 \sqrt{2}A_3A_4+54
A_4^2+108 A_3^2 t^2 -36 \sqrt{2}A_3A_4 t^2-72 A_3^2 t^3\right.\\
&&\left.+48 \sqrt{2}A_3A_4 t^3-24 A_4^2 t^3+36A_3^2 t^4-48
\sqrt{2}A_3A_4 t^4+108 A_4^2 t^4-24 A_4^2 t^5+8A_4^2 t^6+36
\sqrt{2}A_3^2 x\right.\\
&&\left.+108A_3A_4 x+54 \sqrt{2}A_4^2 x-72 \sqrt{2}A_3^2 t^2 x+144
A_3 A_4 t^2 x-36 \sqrt{2}A_4^2 t^2 x-48A_3 A_4 t^3 x\right.\\
&&\left.+48 \sqrt{2}A_4^2 t^3 x+24A_3A_4 t^4 x-48 \sqrt{2}A_4^2 t^4
x+36A_3^2 x^2+54 \sqrt{2}A_3 A_4 x^2+54A_4^2 x^2-36 A_3^2 t
x^2\right.\\
&&\left.+36A_3^2 t^2 x^2-72 \sqrt{2}A_3A_4t^2 x^2+72A_4^2 t^2
x^2-24 A_4^2 t^3 x^2+12 A_4^2 t^4 x^2+36 A_3 A_4 x^3\right.\\
&&\left.-12A_2 \left(A_4 (-3+6 t^2-8 t^3+4 t^4-3 \sqrt{2} x-3
x^2-x^4)+A_3(6 t^2\sqrt{2}-6 t^2x+6 t x-3\sqrt{2}-6
x-3x^3)\right)\right.\\
&&\left.+12A_1 \left(6A_2 x+A_3(6 t-6 t^2+3 x^2)+A_4 (6
t^2\sqrt{2}-6
t^2x+6 t x+x^3)\right)\right.\\
&&\left.+18 \sqrt{2} A_4^2 x^3-24A_3 A_4 t x^3+24A_3A_4 t^2 x^3-24
\sqrt{2} A_4^2 t^2 x^3+9 A_3^2 x^4+9 A_4^2 x^4-6 A_4^2 t x^4\right.\\
&&\left.+6 A_4^2 t^2 x^4+6 A_3A_4 x^5+A_4^2 x^6+36A_2^2 (1-2 t+2
t^2+x^2)\right].
\end{eqnarray*}
\end{figure*}


\begin{thebibliography}{99}
\bibitem{N.Akhmediev} N. Akhmediev, J.M. Soto-Crespo, A. Ankiewicz,
Phys. Lett. A {\bf373}, 2137-2145 (2009).
\bibitem{C. Kharif}C. Kharif and E. Pelinovsky, Eur. J. Mech. B/Fluids {\bf22}, 603
(2003).
\bibitem{D.R. Solli} D.R. Solli, C. Ropers, P. Koonath, B. Jalali, Nature {\bf450}, 06402
(2007); B. Kibler, J. Fatome, C. Finot, G. Millot, et al., Nature
Phys. 6, 790 (2010).
\bibitem{A. Chabchoub} A. Chabchoub, N.P. Hoffmann, and N. Akhmediev, Phys. Rev. Lett.
{\bf106}, 204502 (2011).
\bibitem{Ankiewicz} A. Ankiewicz, J. M. Soto-Crespo, and Nail Akhmediev,
Phys. Rev. E 81, 046602 (2010).
\bibitem{R. Osborne} N. Akhmediev, A.
Ankiewicz, M. Taki, Phys. Lett. A {\bf373}, 675-678 (2009).
\bibitem{V. Voronovich} V. V. Voronovich,
V. I. Shrira, and G. Thomas, J. Fluid Mech. {\bf604}, 263 (2008).
\bibitem{Akhmediev} N. Akhmediev, A. Ankiewicz, and J. M. Soto-Crespo, Phys. Rev. E
{\bf80}, 026601 (2009).
\bibitem{Baronio} F. Baronio, A. Degasperis, M. Conforti, and S. Wabnitz, Phys. Rev. Lett. 109, 044102
(2012).
\bibitem{Ling2} B.L. Guo, L.M. Ling, Chin. Phys. Lett. 28, 110202 (2011).
\bibitem{Bludov} Y.V. Bludov, V.V. Konotop, and N. Akhmediev, Eur. Phys.
J. Special Topics 185, 169 (2010).
\bibitem{zhao2} L.C. Zhao, J. Liu, Joun. Opt. Soc. Am.
B 29, 3119-3127 (2012).
\bibitem{Lakshmanan}  M. Vijayajayanthi,
T. Kanna, and M. Lakshmanan, Phys. Rev. A {\bf77}, 013820 (2008); M.
Vijayajayanthi, T. Kanna and M. Lakshmanan, Europ. Phys. Journ. -
Special Topics {\bf 173}, 57-80 (2009).
\bibitem{Zhao} L.C. Zhao, S.L. He, Phys. Lett. A {\bf375}, 3017-3020
(2011).
\bibitem{Ling}  B.L. Guo, L.M. Ling, Q. P. Liu
, Phys. Rev. E 85, 026607 (2012).
\bibitem{Yang} Y. Ohta and J.K. Yang, Proc. R. Soc. A  468, 1716-1740 (2012).
\bibitem{Cambournac} C. Cambournac, T. Sylvestre, H. Maillotte, B. Vanderlinden,
P. Kockaert, Ph. Emplit, and M. Haelterman, Phys. Rev. Lett. 89, 083901 (2002).
\bibitem{Das} P. Das, T.S.
Raju, U. Roy, and Prasanta K. Panigrahi, Phys. Rev. A {\bf79},
015601 (2009).
\bibitem{Becker} C. Becker, S. Stellmer, P.S. Panahi, S. Dorscher, M. Baumert,
Eva-Maria Richter, Jochen Kronjager, Kai Bongs,Klaus Sengstock,
Nature phys. {\bf4}, 496-501 (2008).
\bibitem{Engels} C. Hamner, J. J. Chang, and P. Engels,
Phys. Rev. Lett. 106, 065302 (2011); M. A. Hoefer, J. J. Chang, C.
Hamner, and P. Engels, Phys. Rev. A 84, 041605(R) (2011).
\bibitem{Erkintalo} M. Erkintalo, K. Hammani, B.
Kibler,C. Finot, N. Akhmediev, J. M. Dudley, and G. Genty, Phys.
Rev. Lett. 107, 253901 (2011).
\end{thebibliography}
\end{document}